\documentclass[11pt,a4paper]{article}
\pdfoutput=1
\usepackage{setspace}
\usepackage{jheppub}
\usepackage{amsfonts,amssymb,epsfig,verbatim,mathbbol,amstext,amsmath,psfrag}
\usepackage{graphicx}
\usepackage[latin2]{inputenc}
\usepackage{t1enc}
\usepackage{amsmath}
\usepackage{subfigure}
\usepackage{dsfont}
\usepackage{slashed}
\usepackage{multirow}
\usepackage{lscape}
\usepackage{wrapfig}
\usepackage{array}
\usepackage{mciteplus}

\usepackage{colordvi}
\usepackage{lscape}
\usepackage{rotfloat}
\usepackage{rotating}
\usepackage{color}

\usepackage{cmap}

\newcommand{\dd}{\textmd{d}}
\newcommand{\be}{\begin{equation}}
\newcommand{\ee}{\end{equation}}
\newcommand{\Tr}{\textmd{Tr}}
\newcommand{\Z}{\mathcal{Z}}

\newcommand{\B}{\mathcal{B}}
\newcommand{\E}{\mathcal{E}}
\renewcommand{\L}{\mathcal{L}}
\newcommand{\D}{\mathcal{D}}

\newcommand{\expv}[1]{\left \langle #1 \right \rangle}

\newcommand{\tr}{\textmd{tr}\,}

\newcommand{\GeVt}{\textmd{ GeV}^2}
\newcommand{\MeV}{\textmd{ MeV}}
\newcommand{\GeV}{\textmd{ GeV}}

\long\def\symbolfootnote[#1]#2{\begingroup%
\def\thefootnote{\fnsymbol{footnote}}\footnote[#1]{#2}\endgroup} 

\title{Critical point in the QCD phase diagram for extremely strong background magnetic fields
}

\author{Gergely~Endr\H{o}di}
\affiliation{Institute for Theoretical Physics, Universit\"at Regensburg, D-93040 Regensburg, Germany}

\emailAdd{gergely.endrodi@physik.uni-r.de}

\abstract{
Lattice simulations have demonstrated that a background (electro)magnetic field
reduces the chiral/deconfinement transition temperature of quantum chromodynamics
for $eB<1\GeVt$. 
On the level of observables, this reduction manifests itself in an enhancement of the Polyakov 
loop and in a suppression of the light quark condensates (inverse magnetic catalysis) 
in the transition region. 
In this paper, we report on lattice 
simulations 
of $1+1+1$-flavor QCD at an unprecedentedly high value of the 
magnetic field $eB=3.25 \GeVt$. 
Based on the behavior of various observables, it is shown that even at this 
extremely strong field,
inverse magnetic catalysis prevails and the transition, albeit becoming sharper, 
remains an analytic crossover. 
In addition, we develop an algorithm to directly simulate the asymptotically 
strong magnetic field limit of QCD. 
We find strong evidence for a first-order 
deconfinement phase transition in this limiting theory, 
implying the presence of a critical point in the 
QCD phase diagram. Based on the available lattice data, we estimate the location 
of the critical point.
}


\begin{document}

\maketitle

\section{Introduction}

Quantum chromodynamics (QCD) exhibits a finite temperature transition 
separating the chirally broken, low-temperature phase from the chirally 
symmetric, high-temperature regime, where quarks and gluons are deconfined.
Although this transition is no real phase transition 
but merely an analytic crossover~\cite{Aoki:2006we,*Bhattacharya:2014ara}, it 
is marked by the pronounced behavior
of the corresponding (approximate) order parameters: the drop in the light quark condensates, 
accompanied by the increase in the Polyakov loop. 
The characteristic dependence of the transition temperature on further parameters 
of the system probes our understanding of QCD and 
maps out the phase diagram in the corresponding 
parameter space. One parameter that is thought 
to have rich physical applications, ranging from neutron star physics through 
heavy-ion collisions to the cosmology of the early universe,
is a background (electro)magnetic field $B$. The relevant range of magnetic 
fields, where QCD interactions compete with the electromagnetic forces, is given 
by multiples of the pion mass squared $m_\pi^2$. 
We refer the reader to recent reviews on the 
subject~\cite{Kharzeev:2013jha,Andersen:2014xxa,Kharzeev:2015kna}.

QCD with background magnetic fields can be studied directly using 
non-perturbative lattice simulations. Continuum extrapolated results 
employing staggered quarks with physical masses 
have been used to map out the 
phase diagram~\cite{Bali:2011qj} for $0\le eB <1\GeVt$. According to these results, the magnetic field 
increases the light quark condensates well below and well above $T_c$ 
(magnetic catalysis) 
but decreases them in the 
transition region~\cite{Bali:2012zg} (inverse magnetic catalysis). 
As a result of this non-monotonous dependence of the condensate 
on $B$ and on $T$, the transition 
temperature is significantly reduced by the magnetic field. 
The same tendency has been observed for the Polyakov loop as well, 
giving a similarly decreasing transition temperature~\cite{Bruckmann:2013oba}. 
More recent lattice simulations employing different quark discretizations 
are consistent with this picture~\cite{Bornyakov:2013eya}.

The magnetic catalysis of the condensate at low temperatures is a very robust concept.
It arises naturally due to the Landau-level structure of charged particle 
energies in the presence of $B$. 
In the strong field limit, magnetic catalysis can be understood in terms of 
the dimensional reduction of the system and the 
high degeneracy of the lowest Landau-level~\cite{Gusynin:1995nb,Schramm:1991ex}.
For low magnetic fields, it can be related to 
the positivity of the QED $\beta$-function that fixes the dependence of 
the condensate on $B$ to order $B^2$~\cite{Endrodi:2013cs,Bali:2014kia}. 
Magnetic catalysis even has connections to solid state 
physics models like the Hofstadter model~\cite{Endrodi:2014vza}. 
In line with these arguments, the catalysis of the condensate 
at low temperatures was observed in a variety of model settings and 
effective theories of QCD. However, in most of these models, magnetic 
catalysis takes place not only for $T<T_c$, but for all temperatures, giving rise to 
a monotonously increasing dependence of $T_c$ on $B$. 
Thus, for the phase diagram, these models predict just the opposite of 
what the above discussed lattice results suggest.
For a recent 
summary on these model approaches and a comparison to the lattice results, see Refs.~\cite{Fraga:2012rr,Andersen:2014xxa}.

While the mechanisms behind magnetic catalysis, as mentioned above, are 
quite transparent,
the opposite behavior around $T_c$ -- inverse magnetic catalysis --
apparently has its 
origin in the rearrangement of the gluonic configurations that dominate the 
QCD path integral and is thus highly nontrivial~\cite{Bruckmann:2013oba}.
Several attempts have been made recently to understand this behavior in 
effective approaches to QCD~\cite{Fukushima:2012kc,*Andersen:2012jf,*Fraga:2012fs,*Fraga:2012ev,*Chao:2013qpa,*Kamikado:2013pya,*Ferreira:2013tba,*Ferrer:2013noa,*Ferrer:2014qka,*Fayazbakhsh:2014mca,*Farias:2014eca,*Ferreira:2014kpa,*Ayala:2014iba,*Ayala:2014gwa,*Andersen:2014oaa,*Mueller:2014tea,*Tawfik:2014hwa,*Tawfik:2015tga}, among others, by introducing new, $B$-dependent model parameters or by taking into account the 
running of the QCD coupling with the magnetic field.
Several of these models exhibit a non-monotonous $T_c(B)$ 
dependence, with an initial reduction followed by an enhancement due to 
the magnetic field. In certain settings, 
it was even shown that no matter how the existing parameters of the model 
are tuned as functions of $B$, the transition temperature always tends to 
rise above a given threshold magnetic field~\cite{Fraga:2013ova}. 

It is just the apparent universality of magnetic catalysis 
that has made the lattice results about inverse catalysis and the 
decreasing $T_c(B)$ dependence 
for $0\le eB < 1\GeVt$ 
so unexpected.
It was speculated 
that magnetic catalysis should reappear at even stronger magnetic fields,
and different hypotheses were recently put forward 
about the strong $B$ regime of the 
phase diagram.
In particular, the strong $B$ limit was argued to induce a new critical point~\cite{Cohen:2013zja}.
The transition temperature was conjectured to turn around and increase 
if the magnetic field is sufficiently strong~\cite{Ilgenfritz:2013ara,Braun:2014fua,Mamo:2015dea,Mueller:2015fka}.
In other cases, $T_c$ was argued to keep decreasing and to hit 
zero~\cite{Kojo:2012js}. 
The transition temperatures for chiral restoration and for deconfinement 
were predicted to split~\cite{Fraga:2008qn} in the presence of the magnetic field, 
and a splitting between the chiral restoration temperature for 
the up and down quarks was also argued to take 
place~\cite{Callebaut:2013ria,Mueller:2015fka}.
Let us refer the reader to the recent reviews~\cite{Fraga:2012rr,Andersen:2014xxa} 
for details. 

In this paper, we aim to check these conjectures by means 
of first-principles lattice simulations 
of $1+1+1$-flavor QCD at an unprecedentedly strong magnetic 
field $eB=3.25 \GeVt$. 
In addition, we also simulate the $B\to\infty$ 
limit directly, 
by considering the effective theory relevant for this limit~\cite{Miransky:2002rp,*Miransky:2015ava}.
We find strong evidence that this limiting theory has a first-order deconfinement phase 
transition and, thus, 
the QCD phase diagram exhibits a critical point 
at strong magnetic fields, where the analytic crossover terminates.
Based on our results, we estimate the location of the critical point, and 
sketch the dependence of the deconfinement transition 
temperature on $B$ over a broad range.
Besides answering a fundamental question about the 
QCD phase diagram, we believe that the results will also be useful for 
building/refining effective theories and models of QCD. 

The paper is organized as follows. In Sec.~\ref{sec:setup} we discuss the details 
of the simulations and define the observables used to study the phase diagram. 
Sec.~\ref{sec:results} contains the lattice results in ordinary QCD, followed by Sec.~\ref{sec:asB}, 
where we discuss the simulations of the anisotropic theory in the asymptotic limit. 
The derivation of this effective theory and the employed
simulation algorithms are discussed in the appendices App.~\ref{app:EH} 
and App.~\ref{sec:aniso_sim}.
Finally, in Sec.~\ref{sec:summary} we summarize our findings regarding the 
QCD phase diagram and conclude.

\section{Setup and observables}
\label{sec:setup}

We consider a spatially symmetric $N_s^3\times N_t$ lattice with 
spacing $a$ so that the temperature is given by $T=(N_ta)^{-1}$, the spatial 
volume by $V=(N_sa)^3$ and the four-volume by $V_4=V/T$. 
Given this geometry, we simulate $1+1+1$-flavor QCD, described by the partition function,
\be
\Z = \int \D U\, e^{-\beta S_g} \!\prod_{f=u,d,s}\![ \det M (U;a^2q_fB, m_fa)]^{1/4},
\label{eq:Z}
\ee
given by the functional integral over the gluonic links $U$. 
We employ stout smeared rooted staggered quarks described by the fermion matrix $M$. 
In Eq.~(\ref{eq:Z}), $S_g\equiv s_g\, V_4$ is the tree-level Symanzik improved gauge action and $\beta=6/g^2$ the inverse 
gauge coupling. For further details of the simulation setup and algorithm, see Refs.~\cite{Aoki:2005vt,Bali:2011qj}.
The parameters of the fermion matrix are the quark masses $m_u=m_d\neq m_s$
and the electric charges $q_u=-2q_d=-2q_s=2e/3$ ($e>0$ is the elementary charge), 
which enter in the product with $B$. The quark masses are set to their 
physical values along the line of constant physics~\cite{Borsanyi:2010cj}.
The magnetic field is oriented along the positive $z$-direction and has 
the quantized flux
\be
\Phi \equiv (aN_s)^2 \,eB = 6\pi N_b, \quad\quad\quad N_b\in \mathds{Z},\quad\quad\quad 0\le N_b<N_s^2,
\ee
where we used that the smallest charge in the system is that of the down quark. 
Lattice discretization effects are suppressed as long as the flux quantum $N_b$ is much smaller 
than the period $N_s^2$. Previous experience suggests that $N_b<N_s^2/16$ is a 
reasonable 
choice~\cite{Bali:2011qj}. 

In the following, we employ the fixed-$N_t$ approach and change the temperature 
$T(\beta)=(N_t \,a(\beta))^{-1}$ by 
varying the inverse gauge coupling. This also implies that a given flux quantum 
corresponds to different magnetic fields at different temperatures,
i.e.\ $eB\propto N_b\,T^2(\beta)$. In particular, we choose $\beta$ values where
a fixed magnetic field $eB=3.25\GeVt$ is represented by integer flux quanta. 
Although this implies that only discrete temperatures are allowed, at this strong 
magnetic field the temperature differences are small enough in order to 
map out the transition region (see below). 

Next, we define the observables that can be used to pin down the transition temperature. 
We begin with the quark condensates and susceptibilities, signaling chiral symmetry,
\be
\expv{\bar\psi\psi_f}\equiv \frac{1}{V_4} \frac{\partial \log\Z}{\partial m_f}, \quad\quad\quad
\expv{\chi_f} \equiv \frac{\partial \expv{\bar\psi\psi_f}}{\partial m_f},
\label{eq:pbpdef}
\ee
and employ the normalization inspired by the Gell-Mann-Oakes-Renner relation, introduced in Ref.~\cite{Bali:2012zg} for the condensate,
\be
\begin{split}
\Sigma_{u,d}(B,T) &= \frac{2m_{ud}}{M_\pi^2 F^2} \left[ \expv{\bar\psi\psi_{u,d}}_{B,T}- \expv{\bar\psi\psi_{u,d}}_{0,0} \right] + 1, \\
\chi_{u,d}^\Sigma(B,T) &= \frac{2m^2_{ud}}{M_\pi^2 F^2} \left[ \expv{\chi_{u,d}}_{B,T}- \expv{\chi_{u,d}}_{0,0} \right].
\end{split}
\label{eq:pbpnorm}
\ee
Here, $M_\pi = 135 \MeV$ is the pion mass and $F=86\MeV$ the chiral limit of the pion 
decay constant.
Both $\Sigma$ and $\chi^\Sigma$ are free of additive and of multiplicative divergences~\cite{Bali:2011qj}. 
In addition, $\Sigma$ is normalized to be unity at $T=B=0$ and approaches zero 
above the transition region~\cite{Bali:2012zg}. 
The vacuum values necessary for the additive renormalization were determined 
in Refs.~\cite{Bali:2011qj,Bali:2012zg}.

The approximate order parameter for center symmetry, related to the 
deconfinement transition, is the Polyakov loop, defined on the lattice as
\be
P=\frac{1}{V} \expv{\sum_\mathbf{x} \Tr\prod_t U_4(t,\mathbf{x})}.
\label{eq:ploop}
\ee
In full QCD, the fermion determinant breaks center symmetry explicitly, so that the 
spontaneous breaking always occurs towards the real center element and $\expv{\textmd{Re}\,P}$ 
is a valid (approximate) order parameter. 
In pure gauge theory (this will be relevant for the $B\to\infty$ limit, see Sec.~\ref{sec:asB}), 
there is no explicit breaking and the three center sectors are equivalent. 
In this case, it is convenient to consider the projection of the Polyakov loop to the nearest center element (see, e.g., Ref.~\cite{Fukugita:1989yw}),
\be
P^{\rm pr} = 
\begin{cases}
\textmd{Re} P, & \arg P \in [-\pi/3,\pi/3], \\
\textmd{Re} [ P e^{-i2\pi/3} ] , & \arg P \in (\pi/3,\pi], \\
\textmd{Re} [P e^{i2\pi/3} ] , & \arg P \in (-\pi,-\pi/3). \\
\end{cases}
\label{eq:pprojdef}
\ee
Simulating pure gauge theory on a finite lattice, $\expv{P}$ always vanishes due to the tunneling between 
center sectors, while $\expv{P^{\rm pr}}$ is positive in the deconfined phase.
The susceptibility of the projected Polyakov loop is defined as
\be
\chi_{P^{\rm pr}} = V \left[ \expv{{P^{\rm pr}}^2} - \expv{P^{\rm pr}}^2 \right].
\label{eq:psuscdef}
\ee

The Polyakov loop renormalizes multiplicatively, with a temperature-dependent 
renormalization constant 
\be
P_r(T,B)=Z(T)\cdot P(T,B)
\ee
which is determined by enforcing $\expv{P_r(T_\star,0)}= P_\star$ and 
we chose $T_\star=162\MeV$ 
and $P_\star=1$. The renormalization was discussed in detail and $Z(T)$ was 
determined in Ref.~\cite{Bruckmann:2013oba}. Notice that while $\textmd{Re}P<3$ by construction, the 
renormalized observable has no upper bound. 

An observable that strongly correlates with $P$ -- and, thus, is 
sensitive to the deconfinement transition -- is the strange quark number susceptibility,
\be
c_2^s = \frac{1}{V_4\cdot T^2} \frac{\partial^2 \log\Z}{\partial \mu_s^2},
\ee
Note that $c_2^s$ contains neither additive nor multiplicative divergences.

Finally, the trace anomaly
\be
I^{(\Phi)} = -\frac{1}{V_4} \left.\frac{\partial \log\Z}{\partial \log a}\right|_\Phi
\label{eq:I}
\ee
can be written as a sum of gluonic, fermionic and magnetic 
contributions~\cite{Bali:2014kia},
\be
I^{(\Phi)}(B,T)= \frac{\partial \beta}{\partial \log a} \expv{s_g} - \sum_f \frac{\partial (m_fa)}{\partial \log a} \expv{\bar\psi\psi_f} + b_1 (eB)^2,
\ee
where $b_1=\sum_f (q_f/e)^2/(4\pi^2)$ is the lowest-order QED $\beta$-function coefficient. 
The magnetic term appears due to electric charge renormalization and stems from 
the counter-term canceling the $B$-dependent 
additive divergence of the thermodynamic potential $\log\Z$~\cite{Bali:2014kia}. Note that this 
term is finite and independent of the regularization, once the continuum limit is taken, see 
discussion in Ref.~\cite{Bali:2014kia}.
Notice furthermore that the derivative in the definition of $I$ is evaluated at fixed magnetic 
flux $\Phi$ and not at fixed magnetic field $eB$ [this is indicated by the superscript $(\Phi)$]. 
The need for distinguishing between the two 
directional derivatives was first discussed in Ref.~\cite{Bali:2013esa} and put into 
practice for the trace anomaly in Ref.~\cite{Bali:2014kia}.

\section{Results at \boldmath $eB=3.25\textmd{\bf \;GeV}^2$}
\label{sec:results}

We extend the previously published data on the light condensates and 
susceptibilities~\cite{Bali:2012zg}, on the Polyakov loop~\cite{Bruckmann:2013oba}, 
on the strange quark number susceptibility~\cite{Bali:2011qj}, 
and on the trace anomaly~\cite{Bali:2013esa,Bali:2014kia} using our new results at $eB=3.25\GeVt$. To achieve this magnetic field strength, a temporal lattice 
extent $N_t=16$ turned out to be necessary. 
These $N_t=16$ lattices are finer than the finite temperature configurations 
used in Refs.~\cite{Bali:2011qj,Bali:2012zg,Bruckmann:2013oba} 
($N_t=6$, $8$ and $10$) to extrapolate to 
the continuum limit. Thus, our results -- although not strictly 
continuum extrapolated -- are expected to lie close to the limit $a\to0$.
We use two spatial lattice sizes 
$32^3\times 16$ and $48^3\times 16$ to control finite size effects. 

\begin{figure}[b]
\centering
\mbox{
\includegraphics[width=8.7cm]{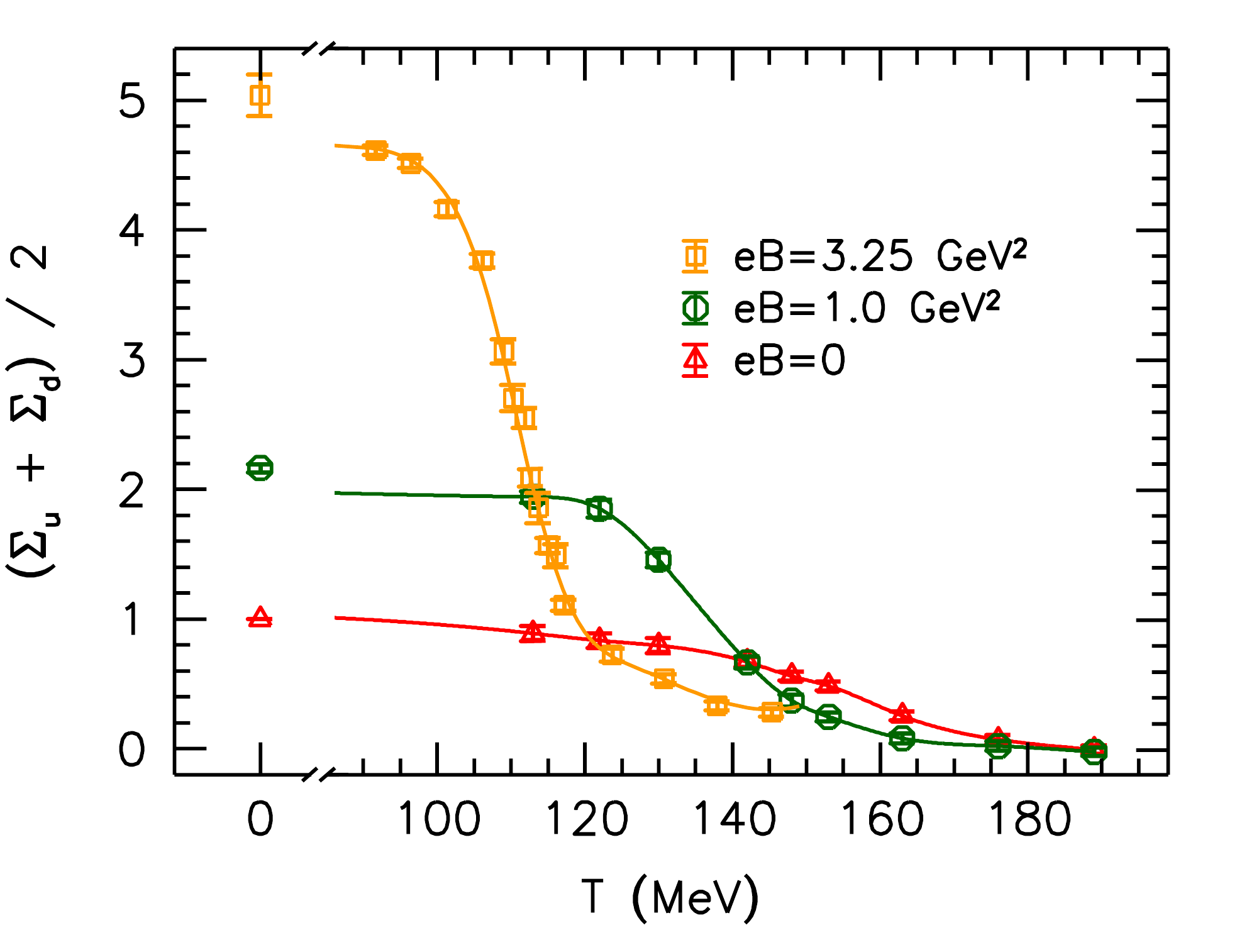}
\includegraphics[width=8.7cm]{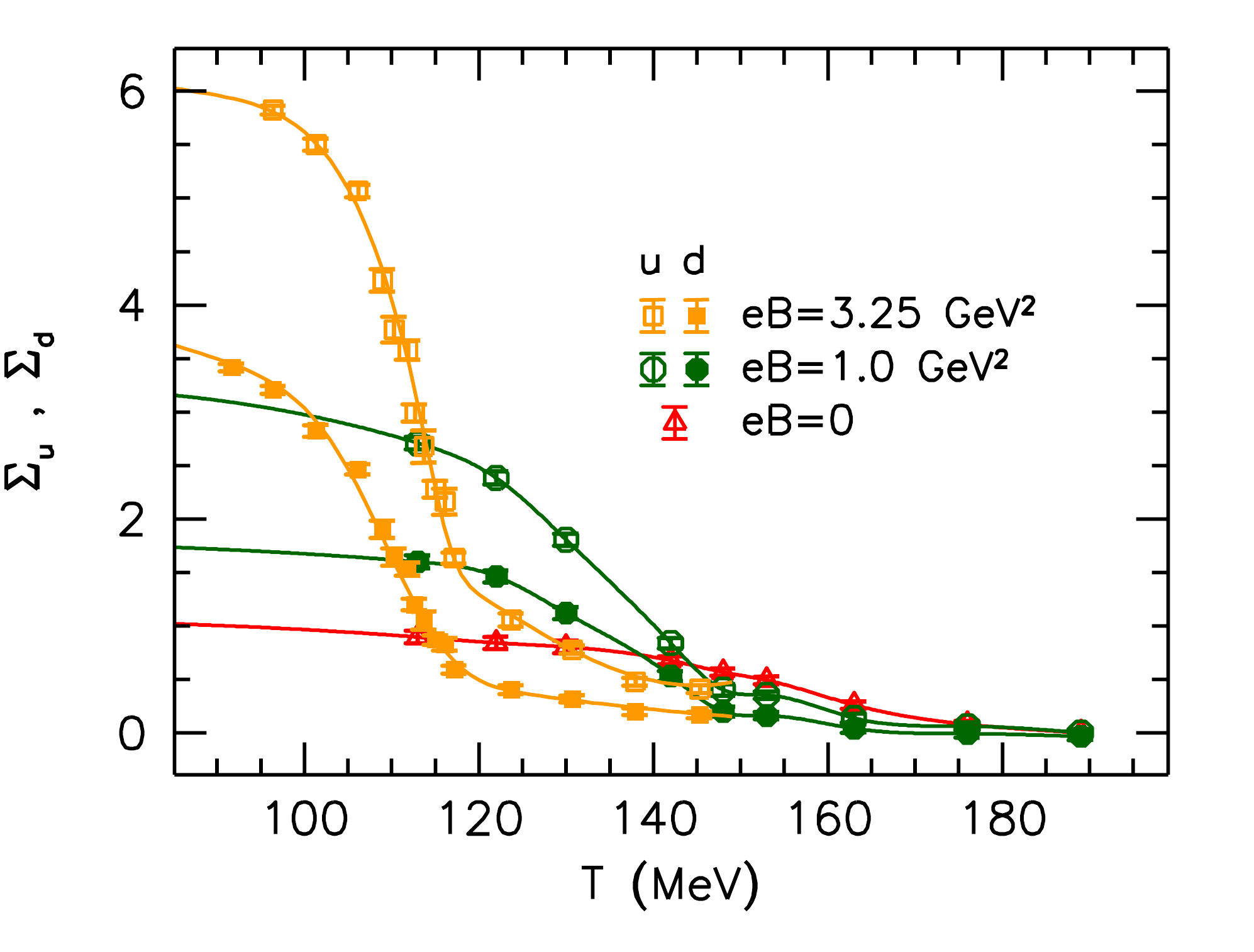}}
\vspace*{-.4cm}
\caption{\label{fig:pbpud} Left panel: average light quark condensate as a function of the temperature 
for three different magnetic fields. 
Right panel: up (open points) and down (filled points) quark condensates for 
the same set of magnetic fields. 
The curves are spline interpolations and merely 
serve to guide the eye.}
\end{figure}

Let us start the discussion with the light quark condensates. The average of $\Sigma_u$ and 
$\Sigma_d$ is plotted in the left panel of 
Fig.~\ref{fig:pbpud}, compared to the $B=0$ and $B=1 \GeVt$ 
continuum extrapolated results~\cite{Bali:2012zg}. 
In addition to the data at nonzero temperatures, we also indicate 
an estimate for the zero-temperature condensate. This is obtained by fitting and 
extrapolating the 
available lattice data 
at $T=0$ by a free-theory inspired form $\sim B\log B$. 
The systematic uncertainty is taken into 
account by varying the fit interval. 

While the condensate is increased by the magnetic field at low temperatures, reflecting 
the well-known magnetic catalysis effect, 
the results also clearly show the reduction of $\Sigma_u+\Sigma_d$ 
in the transition region. Thus, inverse magnetic catalysis is observed to persist in the 
transition region even for 
our strong magnetic field $eB=3.25\GeVt$, pushing $T_c$ further down. In particular, we 
employ the inflection point of the average condensate to find $T_c\{\Sigma_{ud}\}=112(3) \MeV$.
As a side-remark, we mention that since the transition region is shifted to considerably 
lower temperatures, the vacuum values determined for $3.45<\beta<3.85$ in 
Refs.~\cite{Bali:2011qj,Bali:2012zg} suffice to 
perform the additive renormalization of the condensates, and there is no need for 
additional $T=0$ simulations on finer lattices.

Due to the different electric charges, $\Sigma_u$ is expected to be more sensitive to the 
magnetic field than $\Sigma_d$. On that account, even a splitting in the transition temperatures 
might seem plausible, see Refs.~\cite{Callebaut:2013ria,Mueller:2015fka}. 
To check whether this is the case, in the right panel of Fig.~\ref{fig:pbpud} the two condensates are plotted 
separately. Even though the difference $\Sigma_u-\Sigma_d$ is pronounced throughout 
the temperature range in question,
fitting for 
the inflection points gives consistent values $T_c\{\Sigma_u\}=112(3)\MeV$ and 
$T_c\{\Sigma_d\}=111(3) \MeV$. An apparent implication of this finding is that 
the temperature, at which the transition between the chirally broken and restored phases takes place, 
is encoded in the 
gluonic configurations rather than in the operator insertion. 
In lattice language; $T_c$ seems to be a quantity driven predominantly by sea and not by valence effects. 
This also suggests that purely 
gluonic observables would also exhibit similar transition temperatures. 

\begin{figure}[t]
\centering
\mbox{
\includegraphics[width=8.7cm]{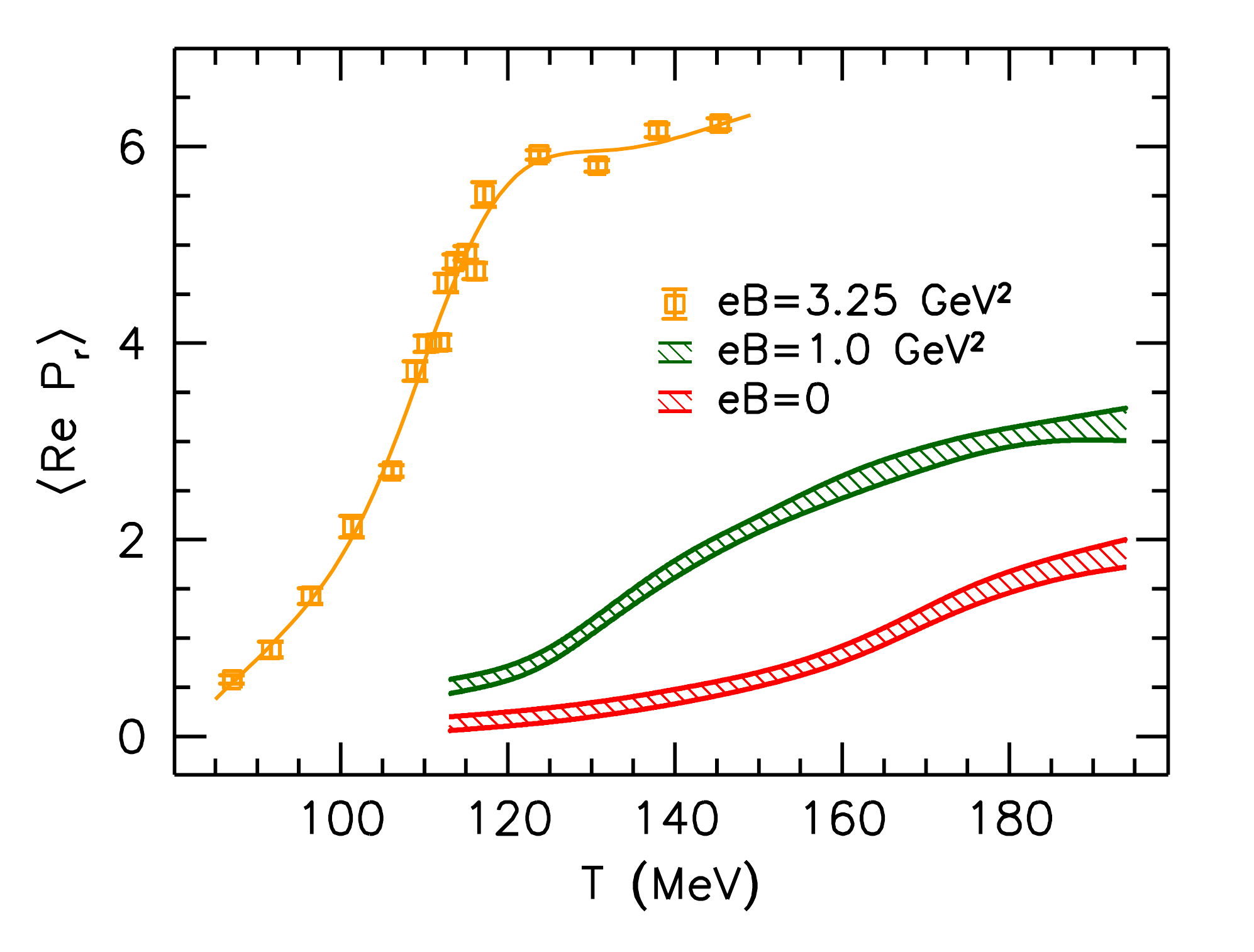}
\includegraphics[width=8.7cm]{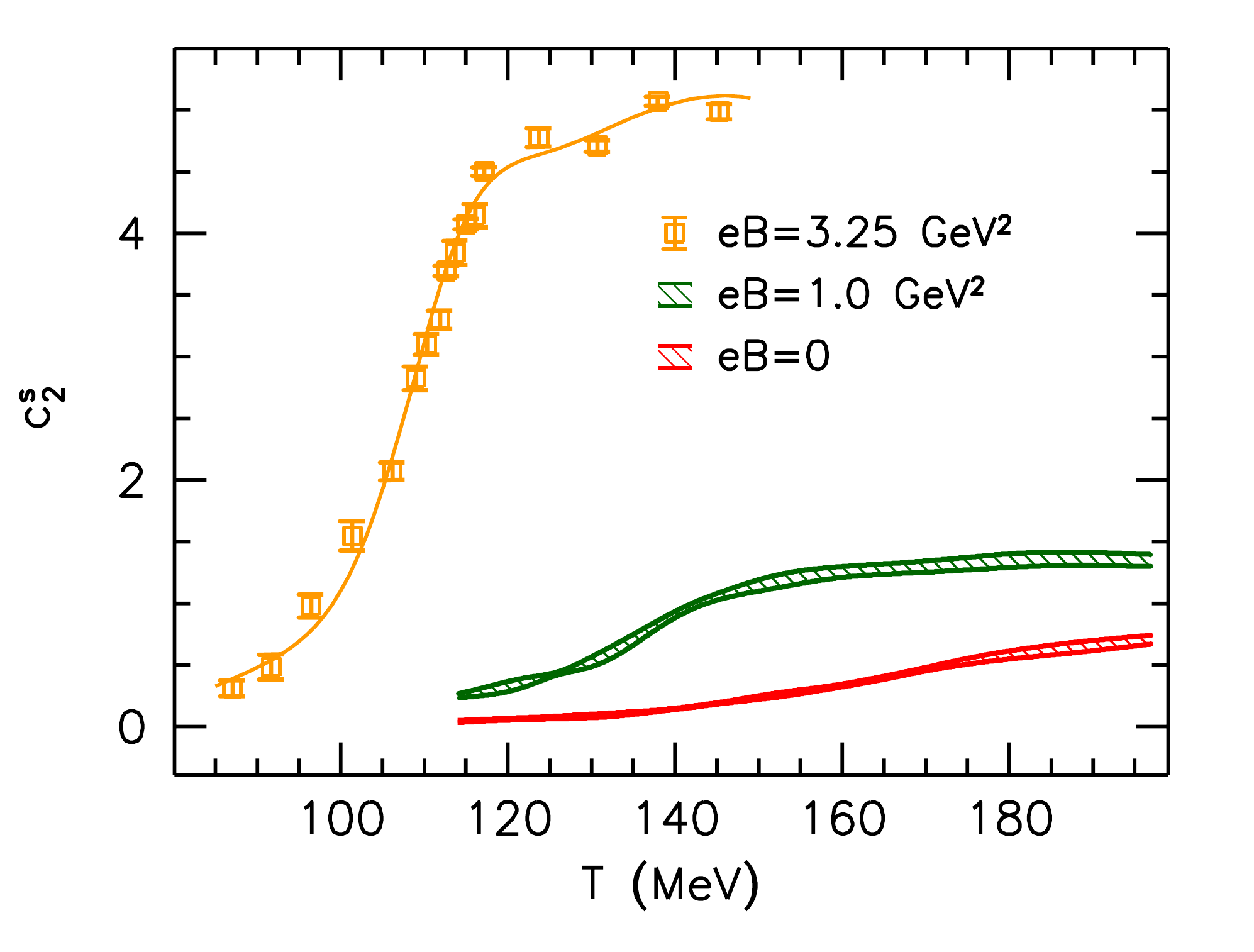}}
\vspace*{-.4cm}
\caption{\label{fig:ploop} Left panel: 
the Polyakov loop for three values of $eB$. At the highest 
magnetic field, the curve is a spline interpolation, while for the lower fields the band 
is the result of a combined continuum extrapolation and interpolation 
in $T$\protect~\cite{Bruckmann:2013oba}.
Right panel: the strange quark number susceptibility for the same set of magnetic fields.
For the highest magnetic field, 
a spline interpolation is shown, whereas for the lower fields the bands represent a 
continuum estimate based on the results of Ref.~\protect\cite{Bali:2011qj}.
}
\end{figure}

This brings us to the simplest, purely gluonic quantity: the Polyakov loop~(\ref{eq:ploop}). 
The (real part of the) renormalized observable is plotted in the left panel of 
Fig.~\ref{fig:ploop}, for the same set of 
magnetic fields, and is observed to be drastically enhanced by the magnetic field for all 
temperatures. The inflection point of $P_r(T)$ is much more pronounced as compared to the 
case at $B=1\GeVt$ and is determined to be $T_c\{P\}=109(3)\MeV$. This value is indeed consistent 
with the transition temperatures obtained above for the light quark condensates. We conclude that 
the gluonic configurations are vastly different on the two `sides' of the transition, and 
predestine the behavior of the light condensates, independently of the electric charge that 
appears in the operator. 
We also observe the strange quark number susceptibility to exhibit
an analogous trend, see the right panel of Fig.~\ref{fig:ploop}. 
Performing a similar fit as for $P_r$, we obtain $T_c\{c_2^s\}=109(3)\MeV$.

A further observable of interest for the QCD equation of state is the trace anomaly~(\ref{eq:I}). 
It measures the breaking of conformal symmetry by the gluonic condensate, by the quark 
condensates and by the magnetic field itself. As $B$ grows, the latter effect becomes 
dominant and $I^{(\Phi)}$ is increased drastically, 
\begin{wrapfigure}{r}{9cm}
\centering
\vspace*{-.1cm}
\includegraphics[width=8.4cm]{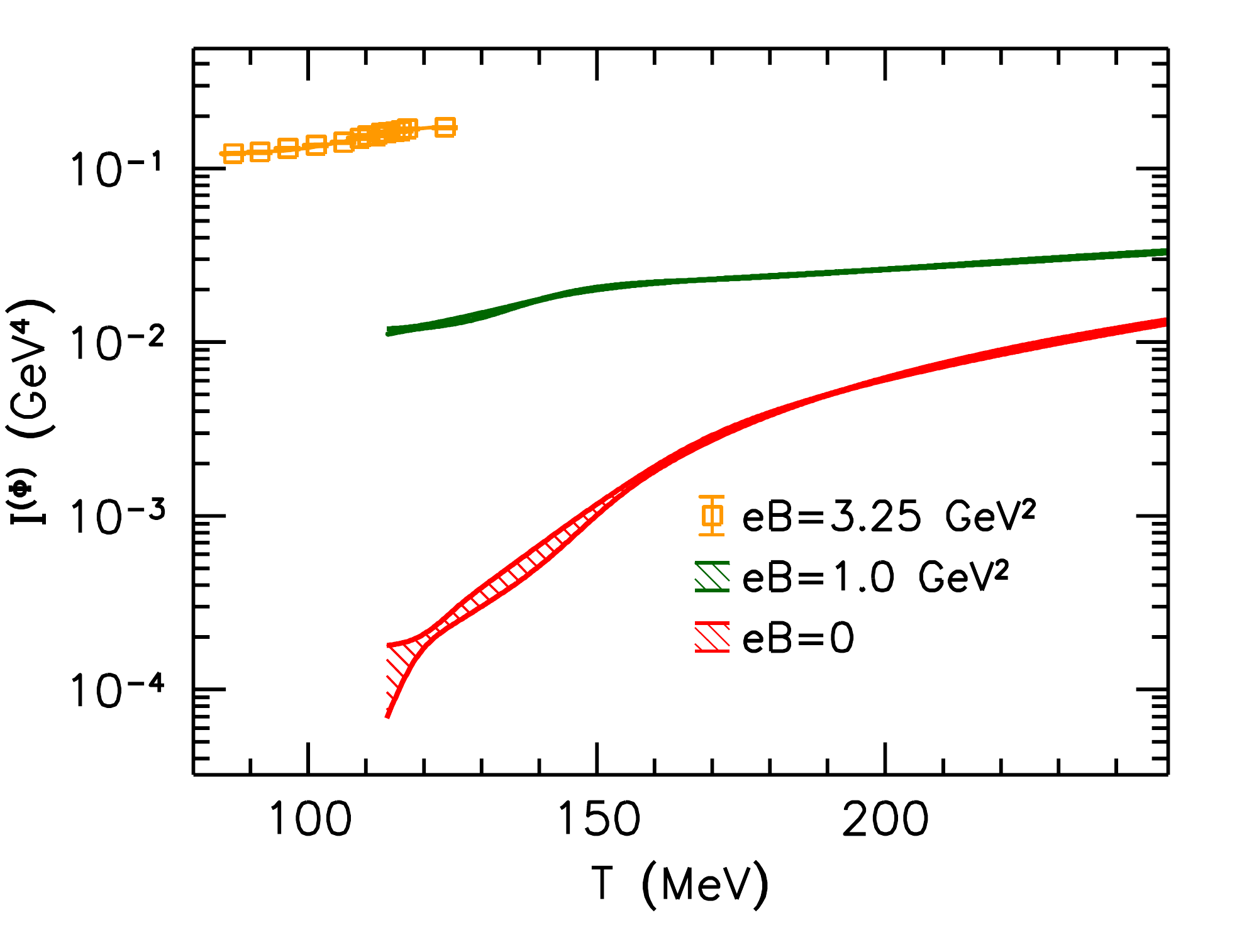}
\vspace*{-.3cm}
\caption{\label{fig:I} The trace anomaly for three different magnetic fields. Note the logarithmic
scale.}
\vspace*{-.2cm}
\end{wrapfigure}
as visible in 
Fig.~\ref{fig:I}. 
Since $I^{(\Phi)}$ contains $B$-dependent contributions already at zero temperature, 
the usual normalization $I^{(\Phi)}/T^4$ is not useful~\cite{Bali:2014kia}. This large 
$T=0$ contribution
also damps the
behavior of $I^{(\Phi)}$ in the transition region. The small kink 
around $T_c$, moving towards smaller temperatures as $B$ grows, is to some extent still visible. 
We note that in order to determine
further quantities related to
the equation of state (e.g.\ pressure, 
entropy density etc.), one would need additional simulations at low temperature 
(see the method developed in Ref.~\cite{Bali:2014kia}). This is outside the scope of the 
present paper. 

\begin{figure}[b]
\centering
\vspace*{-.2cm}
\mbox{
\includegraphics[width=8.4cm]{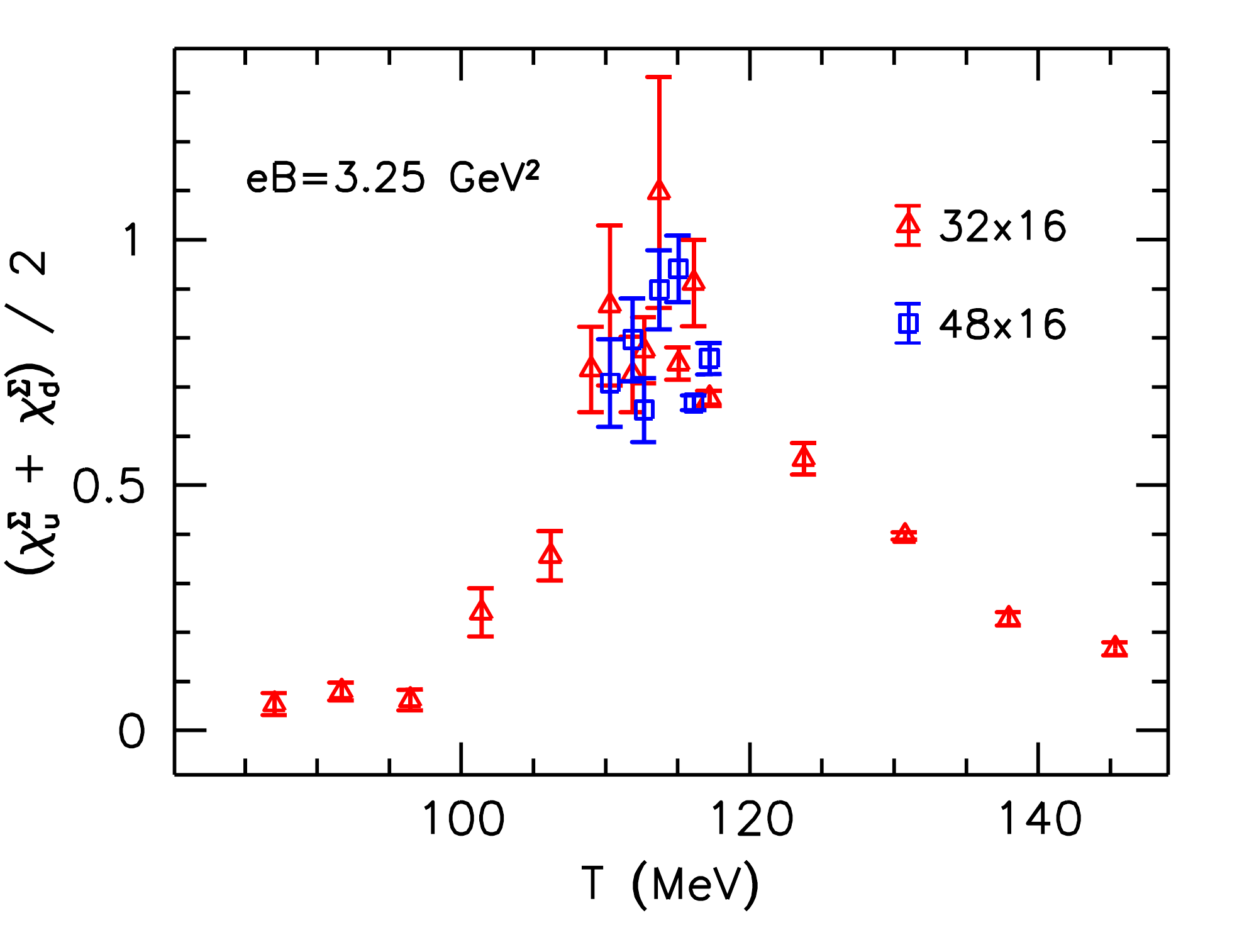}\quad
 \includegraphics[width=8.4cm]{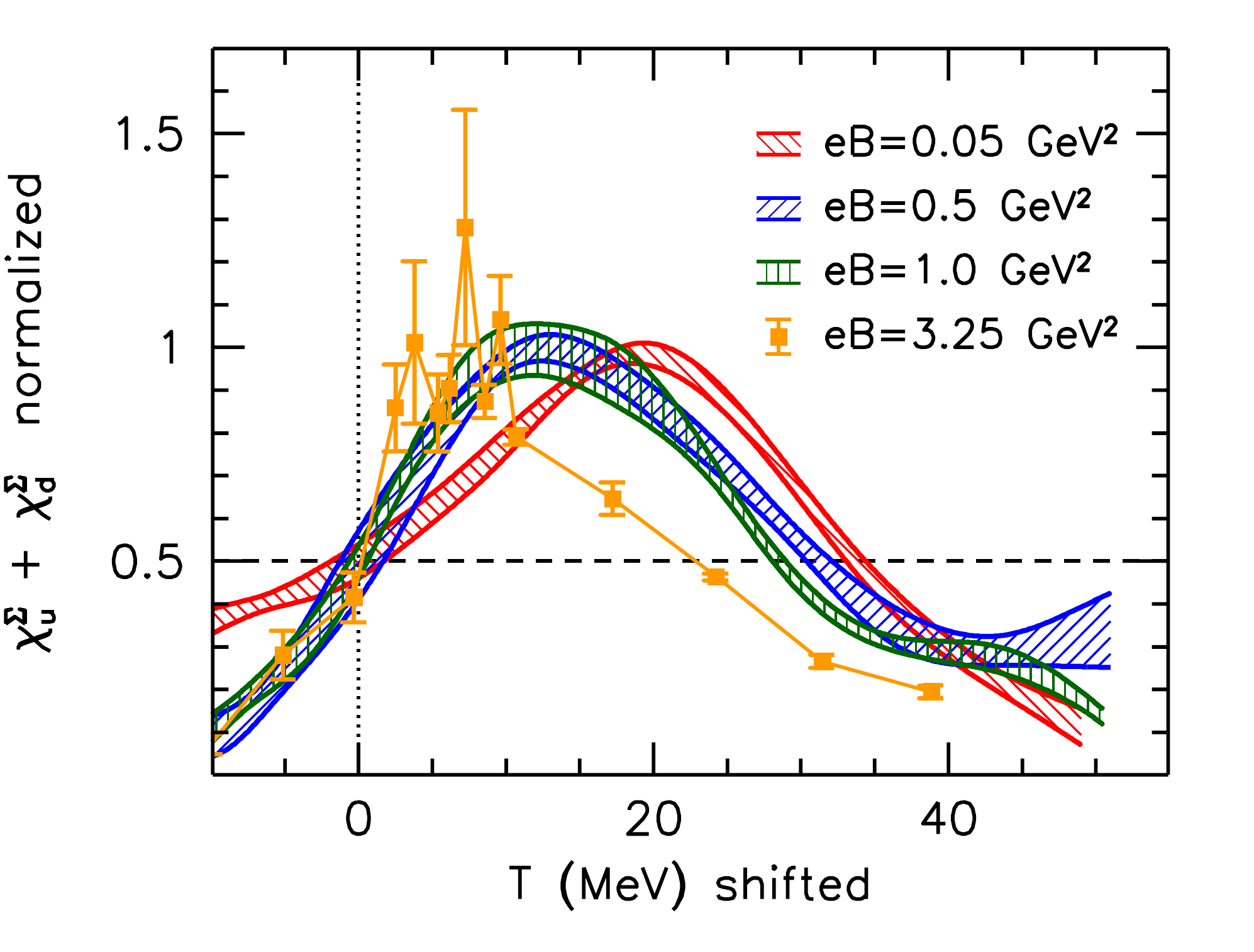}}
\vspace*{-.8cm}
\caption{\label{fig:susc_ud} Left panel: finite size scaling of the average light quark susceptibility.
Right panel: the dependence of the peak width on the magnetic field.}
\end{figure}

Besides the characteristic temperature, the strength of the transition at high magnetic fields 
is also of interest. To determine, whether the smooth crossover at $eB<1\GeVt$ turns 
into a real phase transition at $eB=3.25\GeVt$, we analyze the average of the 
light quark susceptibilities $\chi^{\Sigma}_u+\chi^{\Sigma}_d$. This observable exhibits 
a peak at the transition temperature, see the left panel of 
Fig.~\ref{fig:susc_ud}. For real phase transitions, 
the height $h$ of this peak diverges in the infinite volume limit: 
$h\propto V$ for first-order transitions and $h\propto V^\alpha$ with a critical exponent 
$\alpha<1$ for second-order transitions. In contrast, for the case of an analytic crossover, 
$h$ is independent of the volume. To perform this finite size scaling study, we 
compare the results obtained on the $48^3\times 16$ and on the $32^3\times 16$ ensembles. 
The left panel of Fig.~\ref{fig:susc_ud} 
shows no sign of a singularity as $V$ is increased (note that for a first-order transition, 
the peak heights for the two volumes would differ by more than three). This leads 
us to conclude that the transition remains an analytic crossover even at $eB=3.25\GeVt$. 
We mention moreover that finite volume effects are also absent from the other observables 
discussed above. 

Although the transition remains an analytic crossover, it is instructive to 
analyze the $B$-dependence of the susceptibility peak in more detail. 
We normalize $\chi^\Sigma_u+\chi^\Sigma_d$ such that its peak maximum equals unity,
and plot it in the right panel of Fig.~\ref{fig:susc_ud} against the 
temperature. Here, $T$ is shifted so that the observable equals $0.5$ at zero. 
Then, the peak width $w(B)$ at half maximum can be read off at the 
rightmost intersection of the observable with $0.5$. 
Clearly, $w(B)$ decreases as $B$ grows, signaling that the transition 
becomes stronger in the presence of the magnetic field. We will return to this observation below in 
Sec.~\ref{sec:summary}.

Besides being useful for quantifying the strength of the transition,
the peak of the susceptibility allows for yet another determination of the 
transition temperature. Fitting for the peak maximum, we obtain $T_c\{\chi^\Sigma_{ud}\}=113(4)\MeV$, consistent 
with the results obtained for all other observables. We mention that 
the peak positions for the up and down quark susceptibilities also agree within 
errors. 

\begin{figure}[ht!]
\centering
\includegraphics[width=8.7cm]{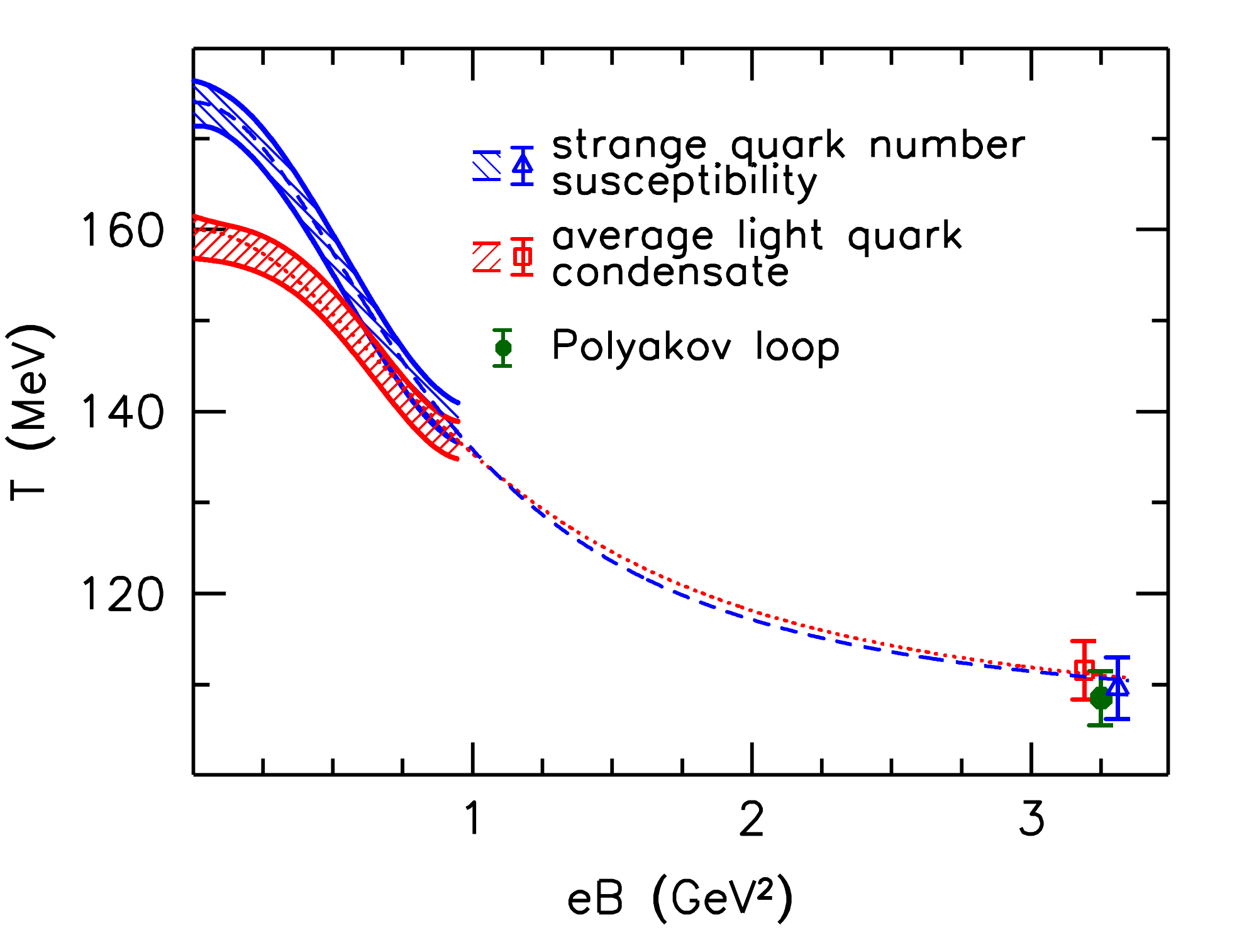}
\vspace*{-.3cm}
\caption{\label{fig:pd} The QCD phase diagram in the magnetic field-temperature plane. 
Previous results at weaker magnetic fields\protect~\cite{Bali:2011qj} are complemented 
by our findings at high $eB$. The points have been slightly shifted horizontally for better 
visibility. The dotted and the dashed lines show an interpolation of the results 
for $\Sigma_u+\Sigma_d$ and for $c_2^s$, respectively, according to Eq.~(\ref{eq:tcfit}).
}
\end{figure}

Finally, in Fig.~\ref{fig:pd} 
we summarize our determinations of $T_c$ in the QCD phase diagram. We consider 
the results for $eB<1\GeVt$ obtained for the light quark condensates and for the 
strange quark number susceptibility~\cite{Bali:2011qj}. In addition, we also include 
the transition temperatures at $eB=3.25\GeVt$ obtained using 
the light quark condensates, the strange quark number susceptibility and 
the Polyakov loop. 
(Note that the inflection point of $P_r$ at $eB<1\GeVt$ is not pronounced enough to 
allow for a stable fit.)
To interpolate $T_c\{\Sigma_{ud}\}$ and $T_c\{c_2^s\}$ for all magnetic fields, we found the following function
sufficient,
\be
T_c(eB) = T_c(0) \cdot \frac{1 + a_1 (eB)^2}{1 + a_2 (eB)^2},
\label{eq:tcfit}
\ee
giving the fit parameters shown in Tab.~\ref{tab:1}. The resulting fit is also shown in the figure.

\begin{table}[ht!]
\setlength{\tabcolsep}{6pt}
\centering
\begin{tabular}{c|c|c|c}
  & $T_c(0)$ & $a_1$ & $a_2$ \\ \hline\hline
 $\Sigma_{ud}$ & 160(2) MeV & 0.54(2) & 0.82(2) \\ \hline
 $c_2^s$ & 174(2) MeV & 0.78(1) & 1.28(1) \\
\end{tabular}
\caption{\label{tab:1} Fit parameters of the function~(\ref{eq:tcfit}).}
\end{table}

\section{Results in the asymptotic magnetic field limit}
\label{sec:asB}

Our results in the right panel of Fig.~\ref{fig:susc_ud} 
indicate that the transition becomes significantly sharper 
as the magnetic field increases. 
This observation raises the question: what happens if $B$ is even larger? 
Does the crossover terminate and turn into a real phase transition? 
To answer this question, we have to consider the limit $eB\gg \Lambda_{\rm QCD}^2$.
Asymptotic freedom dictates that in this limit quarks and gluons 
decouple from each other. Still, the explicit breaking of rotational symmetry by $B$ and 
the corresponding dimensional reduction in the quark sector~\cite{Gusynin:1995nb} 
suggests that this limit is not simply given by a pure gluonic theory plus non-interacting 
(electrically charged) quarks. Indeed, based on the structure of the gluon propagator 
in strong magnetic fields, Ref.~\cite{Miransky:2002rp,*Miransky:2015ava} has shown that the 
effective action describing this limit is an {\it anisotropic} pure gauge theory.
The anisotropy amounts to an enhancement of the chromo-dielectric constant in the 
direction parallel to the magnetic field, characterized by the coefficient $\kappa$,
\be
eB\gg \Lambda_{\rm QCD}^2\,:\quad\quad
\L = \frac{1}{g^2}\left[ \tr \B_\parallel^2 + \tr B_\perp^2 +
(1+g^2\kappa(B))\,\tr \E_\parallel^2 + \tr \E_\perp^2 \right],
\quad\quad
\kappa(B)\propto B.
\label{eq:LagB}
\ee
The definition of the gluonic field strength components 
$\B$ and $\E$ 
is given in Eq.~(\ref{eq:EBdef}) below.
The enhancement of the parallel chromo-dielectric constant implies that the corresponding field strength 
component $\E_\parallel$ is 
suppressed. This tendency is already visible in our full QCD simulations at 
strong magnetic fields, see Fig.~\ref{fig:aniso} in App.~\ref{app:EH} below.

Therefore, as $B$ is increased, the QCD effective Lagrangian 
approaches the anisotropic gauge 
theory given by Eq.~(\ref{eq:LagB}). Assuming
that this theory has a first-order phase transition,
Ref.~\cite{Cohen:2013zja} has conjectured that the strong magnetic field region of the 
QCD phase diagram should exhibit a critical point.
Here we address this question in more detail.
First of all, in App.~\ref{app:EH}, we reproduce the results of Ref.~\cite{Miransky:2002rp,*Miransky:2015ava} 
for the magnetic field-induced anisotropy using the 
effective action in the Schwinger proper-time formulation. 
The resulting anisotropic gauge theory can be simulated directly on the lattice. The setup and the 
simulation algorithm are described in App.~\ref{sec:aniso_sim}. 
The main difference to simple pure gauge theory amounts to multiplying the plaquettes 
lying in the $z-t$ plane by the anisotropy coefficient $\kappa$. The exact form of the 
anisotropic action is given in Eq.~(\ref{eq:Sganiso}). 

Before discussing the lattice simulations of the anisotropic theory, let us 
make one more remark.
Besides writing down the effective Lagrangian~(\ref{eq:LagB}), 
Ref.~\cite{Miransky:2002rp,*Miransky:2015ava} also predicted that the 
scale $\lambda_{\rm QCD}$ 
of this theory 
(generated through dimensional transmutation) 
is much smaller than the QCD scale at $B=0$: 
$\lambda_{\rm QCD} \ll \Lambda_{\rm QCD}$ for a very broad range of magnetic fields. 
In the absence of further dimensionful 
scales in the anisotropic theory, this implies that 
the deconfinement transition temperature for strong mangetic fields is also much smaller than
$T_c(B=0)$. 
Below we will also address this prediction.

\begin{figure}[t]
 \centering
\mbox{
 \includegraphics[width=8.3cm]{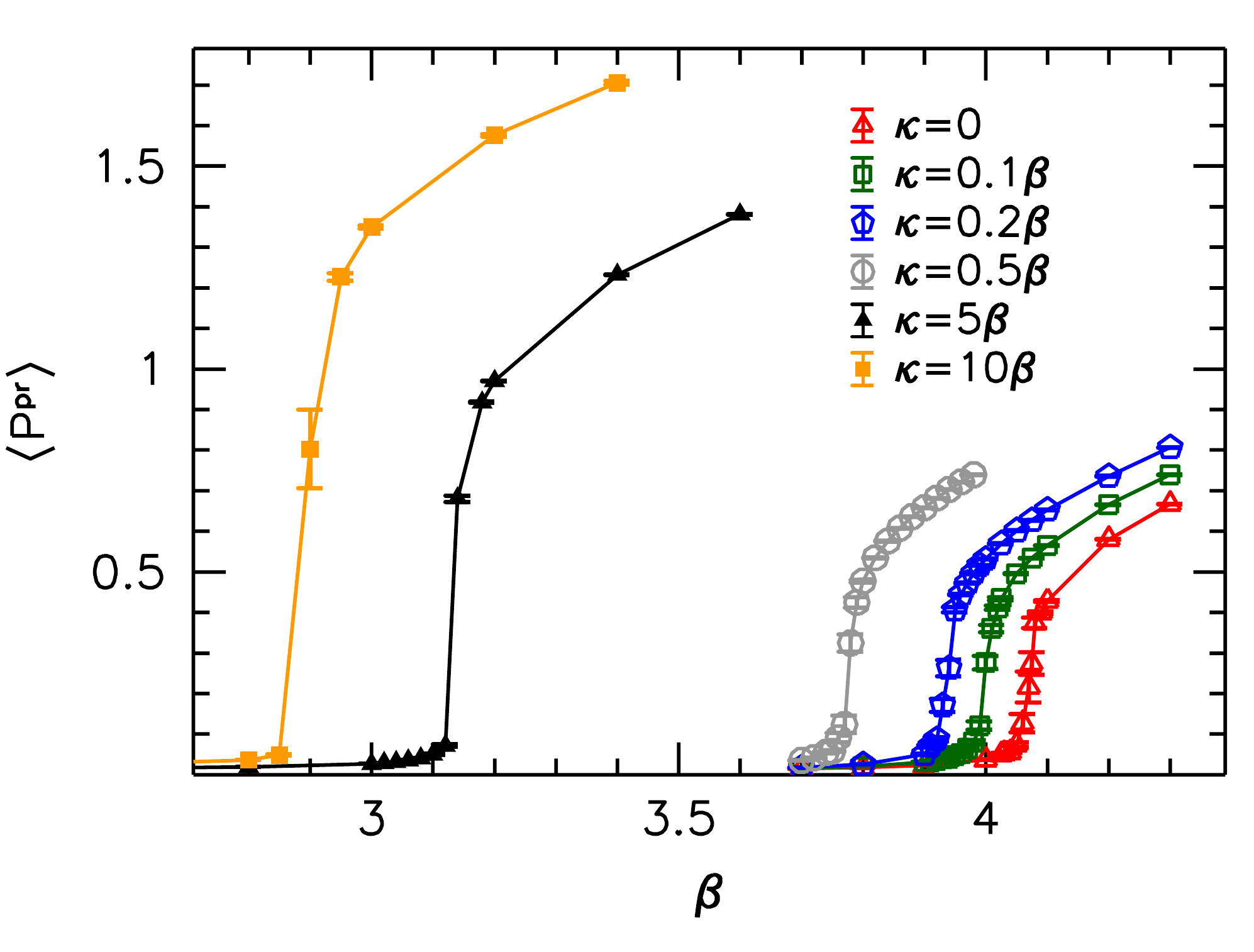} \quad
 \includegraphics[width=8.3cm]{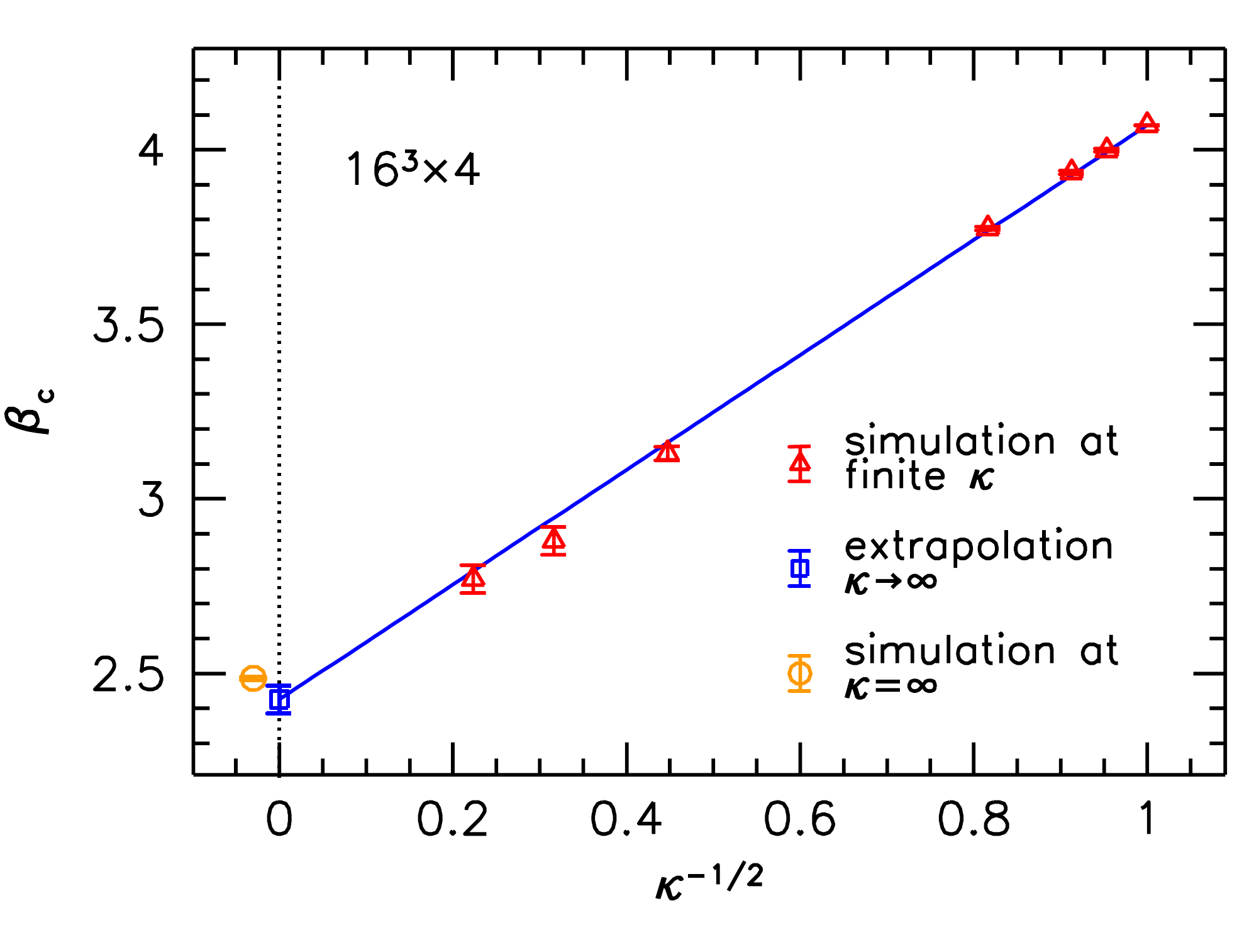} }
 \vspace*{-1cm}
 \caption{\label{fig:kappa1}
   Left panel: the projected Polyakov loop as a function of the inverse gauge coupling 
   for various values of the anisotropy coefficient, as measured on the $16^3\times4$ lattices. The solid lines merely serve to guide the eye. 
   Right panel: the critical inverse coupling as a function of $1/\sqrt{\kappa}$. 
   The extrapolation to $\kappa=\infty$ is compared to the result of the 
   direct simulation at infinite anisotropy (the latter shifted horizontally 
   for better visibility).
 }
\end{figure}

Due to the exact $\mathds{Z}(3)$ symmetry of the anisotropic theory, 
the deconfinement transition is characterized 
by the projected Polyakov loop~(\ref{eq:pprojdef}). 
In the left panel of Fig.~\ref{fig:kappa1}, this observable is plotted
against the inverse gauge coupling $\beta$ for several values of $\kappa$, as measured
on the $16^3\times4$ lattices. 
At $\kappa=0$, we reproduce the results of Refs.~\cite{Cella:1994sx,Borsanyi:2012ve} -- 
in particular, the deconfinement transition occurs at $\beta_c\approx 4.07$. 
The results indicate $\beta_c$ to be strongly reduced as $\kappa$ 
grows\footnote{Here we simulated at fixed values of the ratio $\kappa/\beta$. This continuous 
rescaling of $\kappa$ has no effect on, for example, the critical inverse coupling.}.
 We find that $\beta_c$ scales approximately with $1/\sqrt{\kappa}$, see the right panel 
 of Fig.~\ref{fig:kappa1}. Extrapolating to $\kappa=\infty$ we obtain $\beta_c(\kappa=\infty)\approx 2.42(5)$. 
Besides approaching this limit via finite values of the anisotropy coefficient, 
we also develop an algorithm to simulate directly at $\kappa=\infty$. 
The corresponding setup is described in App.~\ref{sec:aniso_sim}.

\begin{figure}[ht!]
 \centering
 \mbox{
 \includegraphics[width=8.4cm]{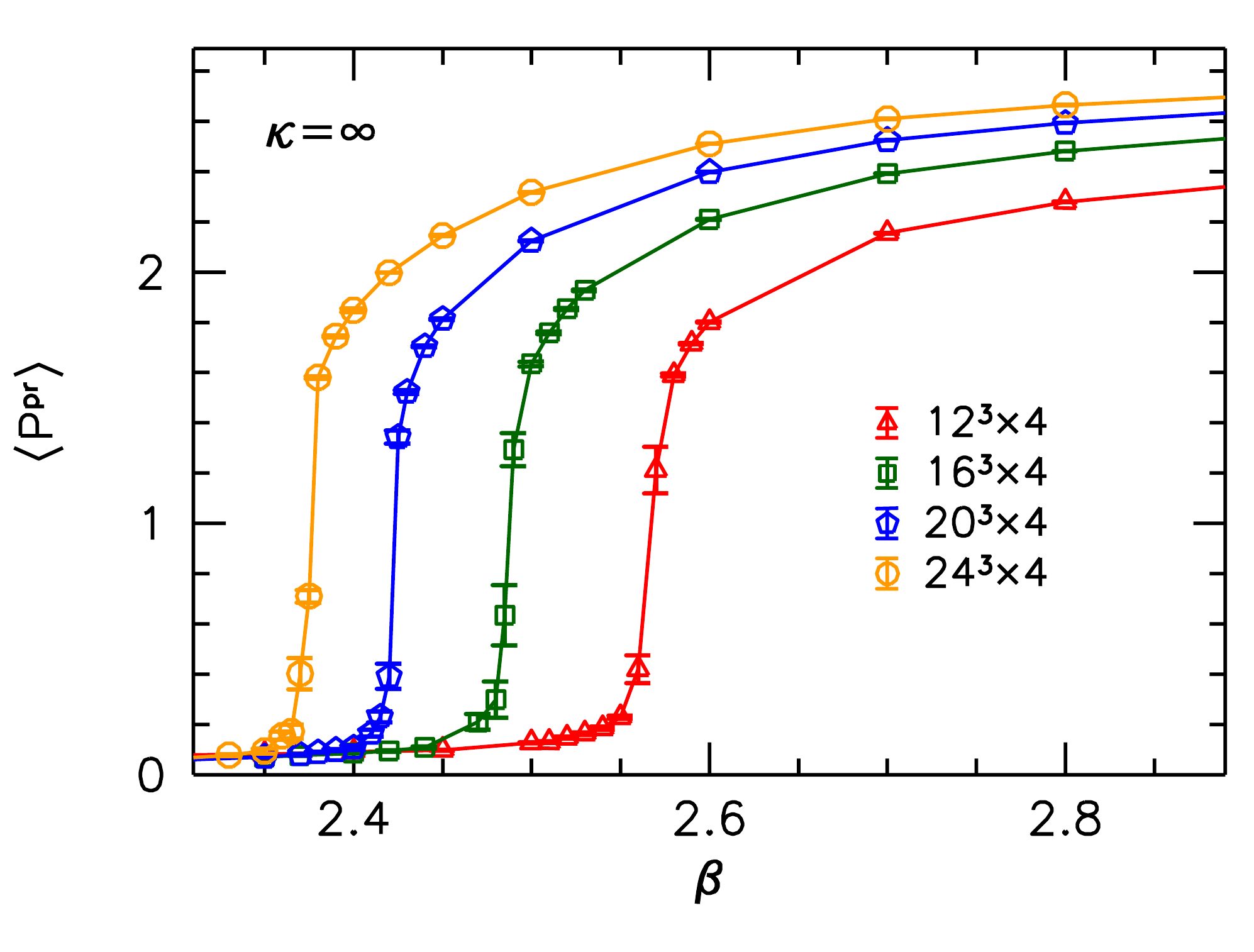}
 \includegraphics[width=8.4cm]{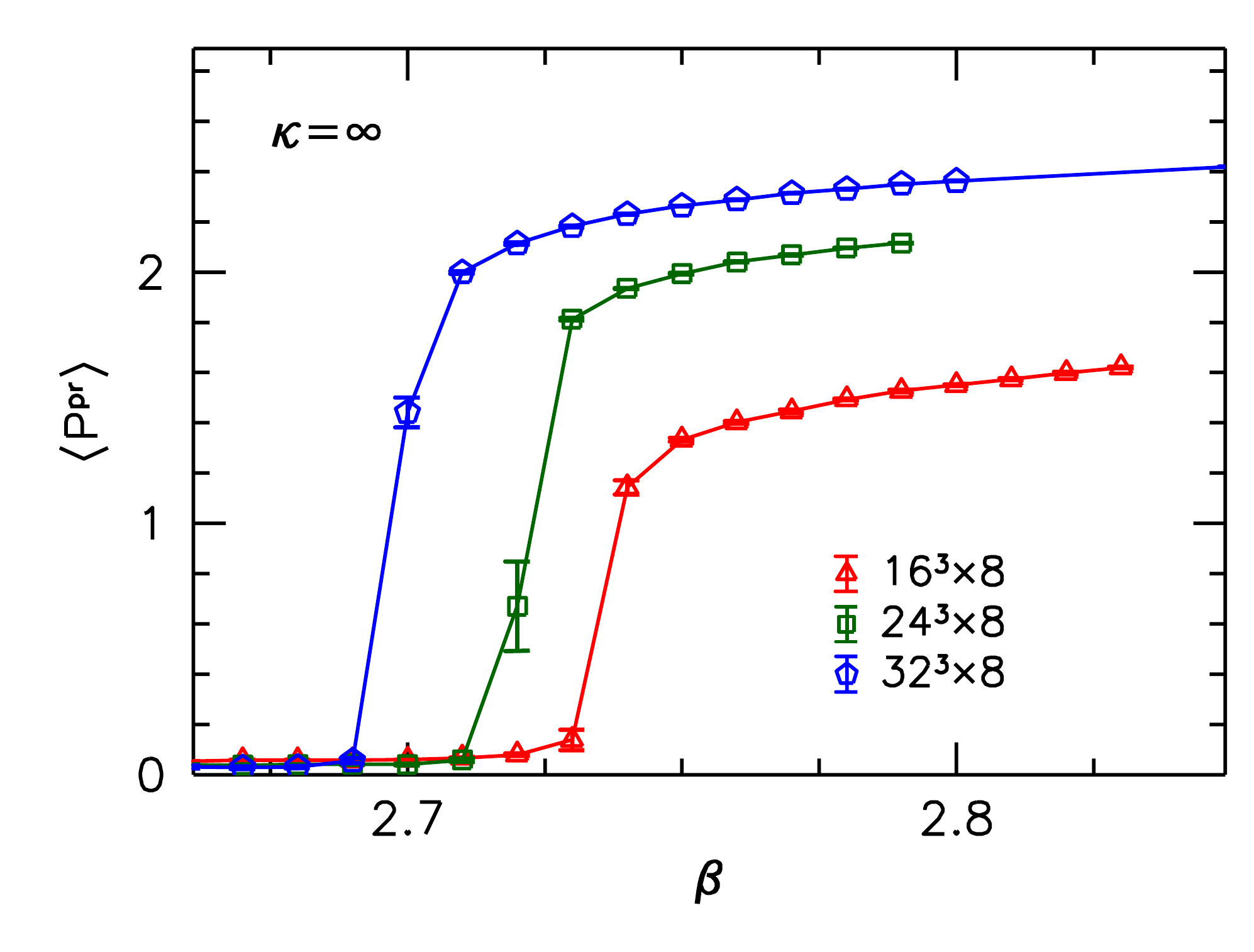} }
 \vspace*{-1cm}
 \caption{\label{fig:kappa2}
   The projected Polyakov loop as a function of the inverse gauge coupling 
   at $\kappa=\infty$, for various lattice volumes with $N_t=4$ (left panel) 
and $N_t=8$ (right panel). The solid lines merely serve to guide the eye.
 }
\end{figure}

The left panel of Fig.~\ref{fig:kappa2} shows $P^{\rm pr}$ at infinite anisotropy. 
We find that the critical inverse coupling on the $16^3\times4$ lattice is 
comparable with the extrapolation based on finite anisotropies, see 
the right panel of Fig.~\ref{fig:kappa1}.
In addition, a 
comparison of the results at different spatial volumes $12^3\ldots 24^3$ reveals 
that the transition 
becomes sharper as the volume increases, as typical for real phase transitions. 
We also repeated this analysis on $N_t=8$ lattices, see the right panel of 
Fig.~\ref{fig:kappa2}. The critical couplings are clearly 
different, showing that the transition is indeed related to the finite temperature.
We also mention that 
finite volume effects in $\beta_c$ are observed to be unusually large -- 
above $10\%$ for $N_t=4$. 
(For comparison, the finite volume 
effects at $\kappa=0$ on similar lattices are of $0.1\%$~\cite{Cella:1994sx}.)
We suspect that this is due to lattice artefacts -- indeed, the effect is 
considerably 
smaller for $N_t=8$, see the right panel of Fig.~\ref{fig:kappa2}. 

\begin{figure}[ht!]
 \centering
 \mbox{
 \includegraphics[width=8.3cm]{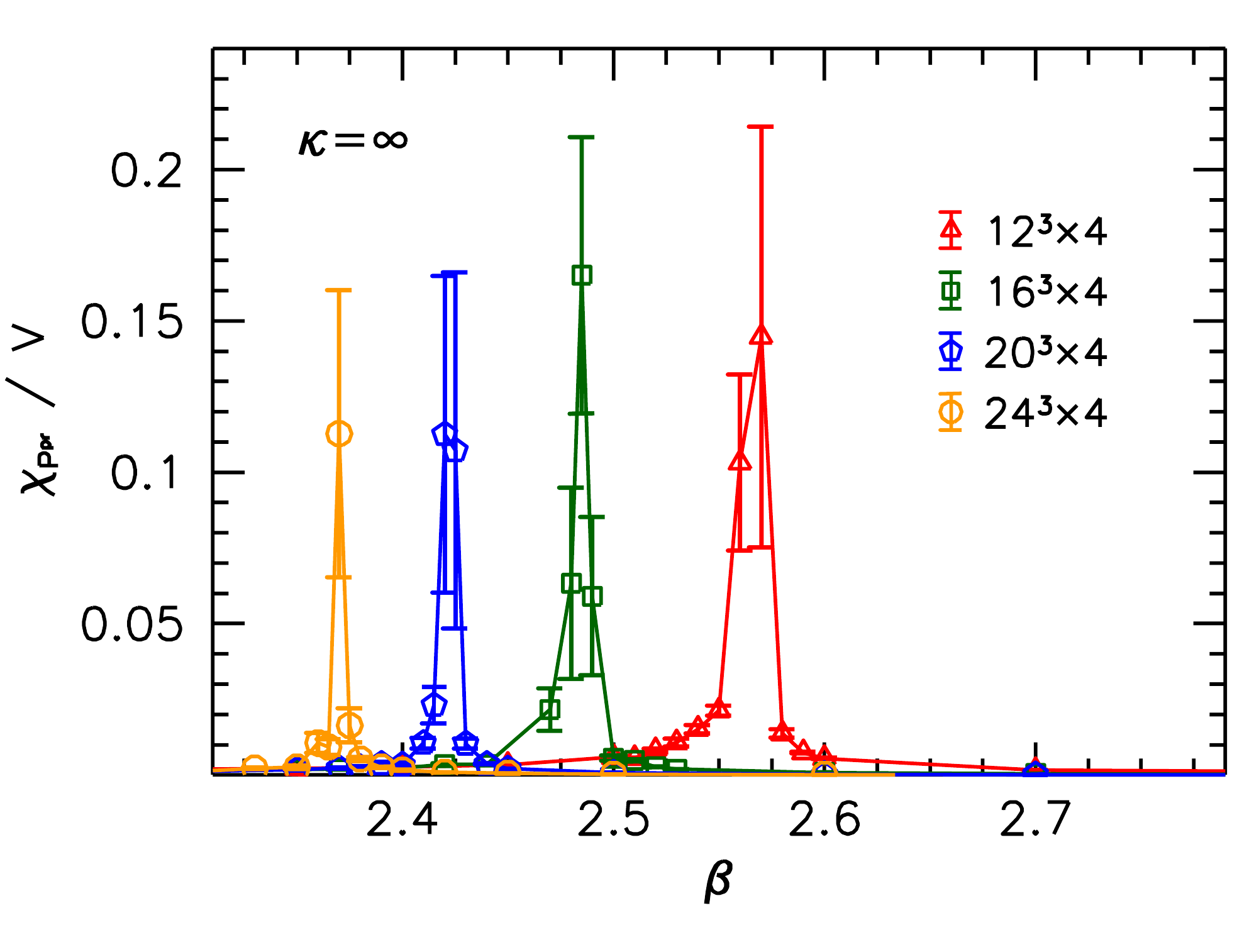} \;
 \includegraphics[width=8.3cm]{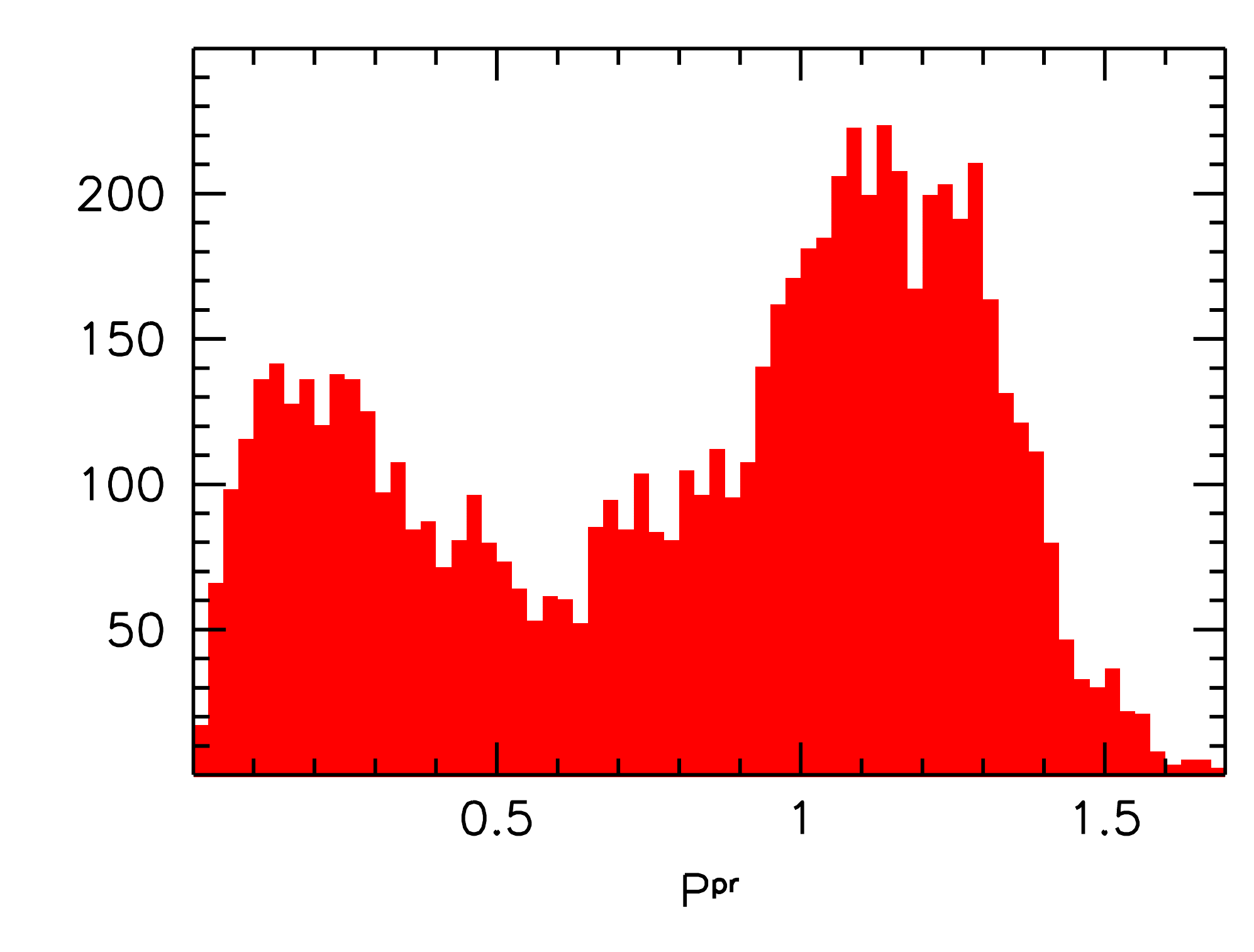} }
 \vspace*{-.8cm}
 \caption{\label{fig:kappa3}
   Left panel: the susceptibility of the projected Polyakov loop, normalized by the spatial volume, 
as a function of the inverse coupling. Various spatial volumes with $N_t=4$ are 
compared. The solid lines serve to guide the eye. 
Right panel: histogram of $P^{\rm pr}$ near the critical temperature 
on the $16^3\times4$ lattices. 
 }
\end{figure}

To determine the nature of the transition, we calculated 
the susceptibility of the projected Polyakov loop~(\ref{eq:psuscdef}). 
This observable is shown in the left panel of 
Fig.~\ref{fig:kappa3}, with a normalization by the spatial 
volume. Within statistical errors, the height of the normalized susceptibility peak 
is observed to be independent of $V$. In other words, the peak height scales linearly 
with $V$, which we take as strong 
evidence that the transition is of {\it first order}. 
The histogram of $P^{\rm pr}$ at $\beta=2.4855$ as measured on the $16^3\times4$ 
lattices is shown in the right panel of Fig.~\ref{fig:kappa3}, revealing the 
two-peak structure characteristic for first-order transitions. 

Through the equivalence between this anisotropic gauge theory and 
QCD with asymptotically strong magnetic fields, the above finding implies that 
the QCD phase diagram exhibits a critical point in the strong magnetic field region, 
where the crossover turns into a real phase transition. 
Based on our full QCD results for the light quark 
susceptibilities, 
we will estimate the magnetic field corresponding to the critical point 
in Sec.~\ref{sec:summary}.

The next step is to relate the critical inverse coupling $\beta_c$ 
to the critical temperature $T_c$ in physical units. 
To do so, we must set the lattice scale $\beta(a)$. 
In principle, the magnetic field is not expected to change this scaling relation 
(cf.\ Ref.~\cite{Bali:2011qj}). However, 
to arrive at our anisotropic gauge theory, $B$ has been taken to infinity, i.e.\ it 
also exceeds the squared lattice cutoff $a^{-2}$. 
Clearly, the lattice scale determined at $B=0$ 
becomes invalid beyond this point. 
Thus, in order to determine the lattice spacing, one needs a dimensionful quantity 
whose value is known in the asymptotic limit -- for example a purely gluonic observable, 
where the $B$-dependence is expected to be only mild.
A possible candidate for this role is the parameter $w_0$
defined from the 
gradient flow of the gauge links~\cite{Luscher:2010iy} that is often used 
for scale setting in QCD, as suggested in Ref.~\cite{Borsanyi:2012zs}. 

\begin{figure}[ht!]
 \centering
 \mbox{
 \includegraphics[width=8.4cm]{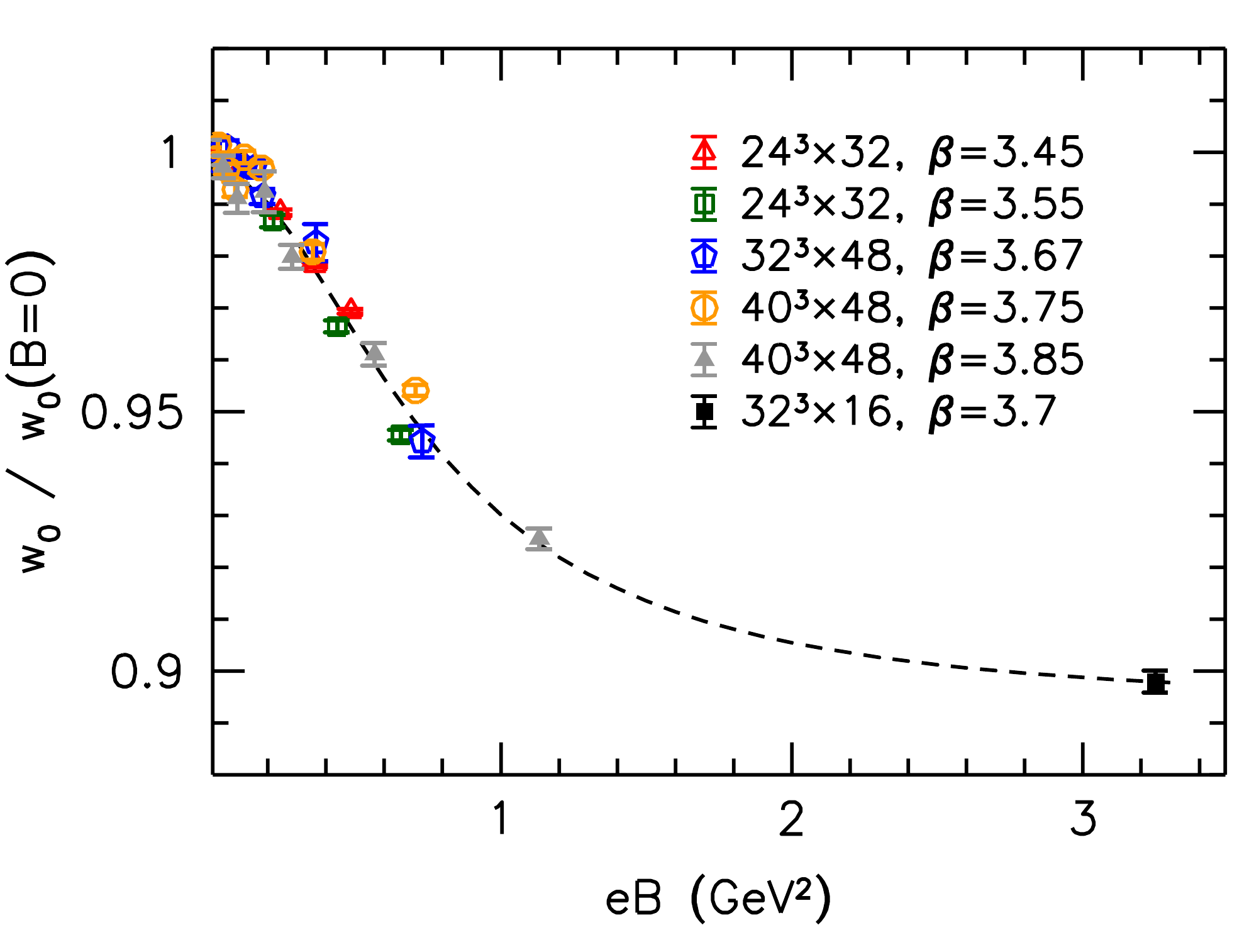}
 \includegraphics[width=8.4cm]{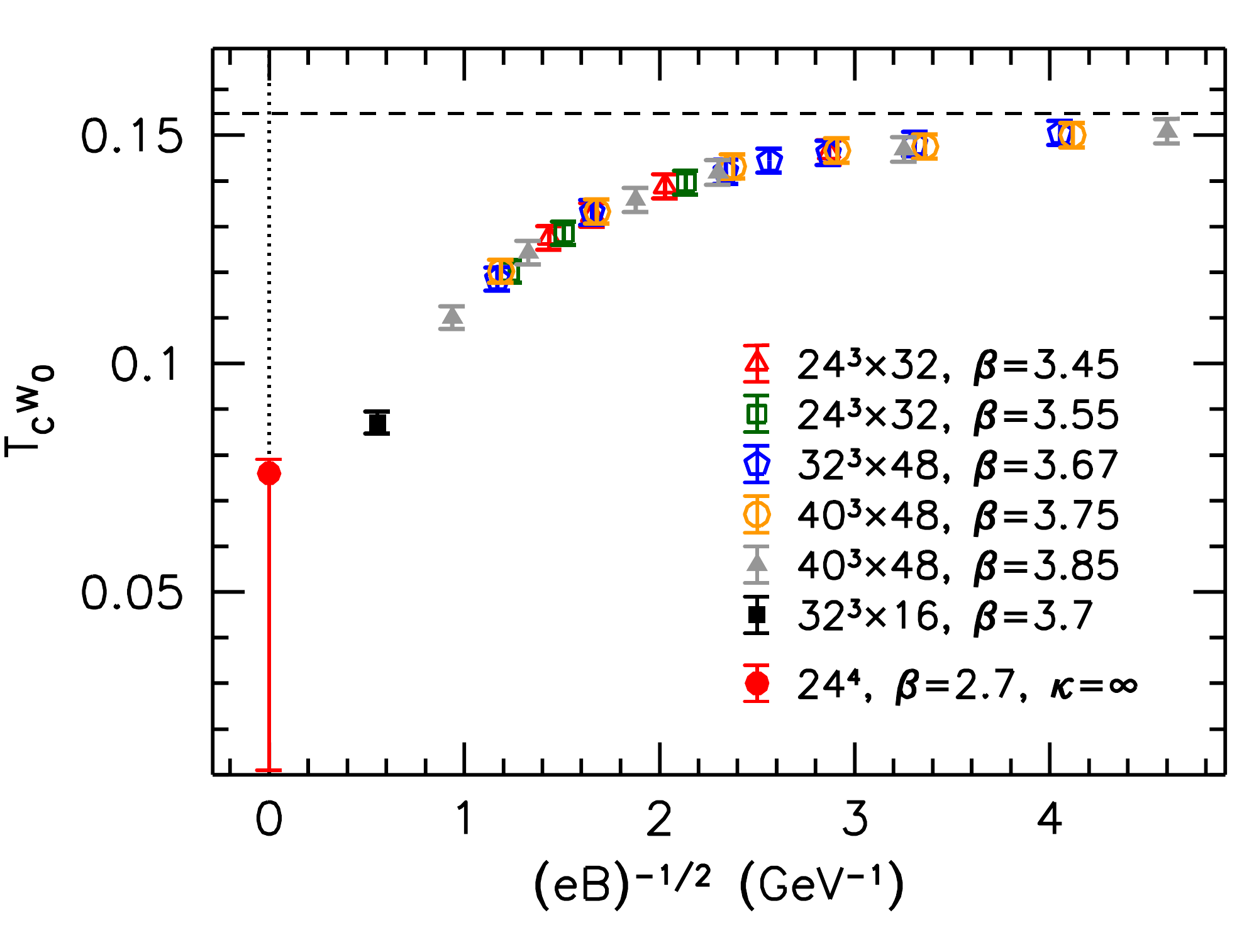}
 }
 \vspace*{-.8cm}
 \caption{\label{fig:w0} Left panel: magnetic field-dependence of 
 the parameter $w_0$ using various lattice spacings and a fit (dashed line) of the 
 form similar to Eq.~(\protect\ref{eq:tcfit}). Right panel: 
 the dimensionless combination $T_cw_0$ in full QCD ($1/\sqrt{eB}>0.5 \textmd{ GeV}^{-1}$)
 and 
 in the anisotropic pure gauge theory ($1/\sqrt{eB}=0$). The dashed line indicates 
 the $B=0$ limit.}
\end{figure}

We determined $w_0$ on our zero-temperature 
full QCD ensembles~\cite{Bali:2012zg} for $eB<1\GeVt$ and also for $eB=3.25\GeVt$ at our 
lowest temperature $T\approx 75\MeV$. The results for the 
ratio $w_0/w_0(B=0)$ are plotted in the left panel of 
Fig.~\ref{fig:w0}, showing a mild reduction of this parameter 
as $B$ grows. A fit of the form similar to Eq.~(\ref{eq:tcfit}) 
describes the data well and suggests a saturation towards the asymptotically 
strong magnetic field limit. Nevertheless, we cannot exclude a significant 
dependence of $w_0$ on $B$ for $B>3.25\GeVt$. 

In addition, we can also gain some insight by considering the 
dimensionless combination $T_cw_0$. 
How close full QCD at $eB=3.25\GeVt$ is to the 
asymptotic limit can then be quantified by matching $T_cw_0$ with the anisotropic 
theory.
Multiplying our full QCD results for $w_0$ by the transition temperature 
(here we take the definition of $T_c$ employing the 
inflection point of the strange quark number susceptibility, cf.\ Fig.~\ref{fig:pd}), 
$T_cw_0$ is shown in the right panel of Fig.~\ref{fig:w0}. Motivated 
by the scaling of $\beta_c$ (cf.\ the right panel of Fig.~\ref{fig:kappa1}), 
the results are plotted against $1/\sqrt{eB}$.
Employing the result for $w_0$ from Ref.~\cite{Borsanyi:2012zs}, 
at zero magnetic field we have $T_c(B=0)\cdot w_0(B=0) = 0.174(3) \GeV\cdot 
0.1755(19)\textmd{ fm} = 0.155(3)$. 

To carry out the comparison to the asymptotic limit, 
we also determined $w_0/a$ on symmetric $16^4$ and $24^4$ 
anisotropic gauge configurations at the critical couplings corresponding 
to the $16^3\times 4$ ($\beta_c\approx 2.47$) and to the $24^3\times8$ 
($\beta_c\approx 2.7$) lattices\footnote{
Just as in full QCD, the gauge links are evolved here using the symmetric gradient flow. 
}.
We observe that the combination $T_cw_0=w_0(\beta_c)/a\cdot 1/N_t$ -- similarly to $\beta_c$ -- 
suffers from large lattice 
discretization effects and exhibits a downwards trend towards the continuum limit. 
We take the result for the $N_t=8$ data 
as an upper limit, 
giving $\lim_{B\to\infty}(T_cw_0) \lesssim 0.076$. 
This value is also included 
in the right panel of Fig.~\ref{fig:w0}. 
Altogether, the results are compatible with 
a monotonous dependence of $T_cw_0$ on $B$. 
To extrapolate $T_cw_0$ reliably to the continuum limit in the anisotropic theory 
requires further simulations 
on finer lattices and will be discussed in a forthcoming publication.

To summarize, the lattice results favor 
a saturation of $w_0$ and a monotonous reduction of $T_cw_0$ 
as the limit $B\to\infty$ is approached. This suggests a monotonous reduction of 
$T_c(B)$ towards the asymptotic limit. 
Nevertheless, based on the available findings, no final statement about 
$\lim_{B\to\infty}T_c$ can be made.

Let us make one more remark about the $\kappa=\infty$ anisotropic theory. Since 
the parallel chromoelectric component $\tr\E_\parallel^2$ of the action vanishes, 
all plaquettes lying in the $z-t$ plane are unity. This implies that all 
Wilson loops $W$ in this plane are trivial, and the static 
quark-antiquark potential $\propto \log W$ is independent of the distance.
Accordingly, 
there is no force acting on quark-antiquark pairs if they are separated in the 
direction of the magnetic field, i.e.\ the string tension $\sigma_{\parallel}$ 
in this direction vanishes. This is in line with recent lattice determinations 
of the string tension in magnetic fields~\cite{Bonati:2014ksa}.

\section{Conclusions}
\label{sec:summary}

In this paper, we determined the nature and the characteristic temperature of the chiral/deconfinement 
transition of QCD at an extremely strong background magnetic field $eB=3.25\GeVt$. 
The results for various observables consistently show that the transition 
temperature is further decreased compared to its value at lower magnetic fields.
For the light quark condensates, the reduction of $T_c$ is due to 
the so-called inverse magnetic catalysis: between $eB=1\GeVt$ and 
$eB=3.25\GeVt$, $\Sigma_{u}$ and of $\Sigma_d$ are 
significantly reduced in the transition region. 
At the same time, the condensates are enhanced by $B$ both for $T\ll T_c$ and 
for $T\gg T_c$ (the latter effect is small, since the condensate 
is suppressed at high temperatures).

Comparing the behavior of the up and down quark condensates and that of the Polyakov loop 
also revealed that there is no splitting between the transition temperatures for the individual 
flavors, neither is there significant difference between the chiral and the deconfinement 
transition temperatures. On the contrary, the different definitions of $T_c$ tend to 
approach each other as $B$ grows and at $eB=3.25\GeVt$ all observables exhibit 
a single transition temperature of around $109-112 \MeV$, see Fig.~\ref{fig:pd}. 
Furthermore, we performed a finite size scaling analysis of the light quark susceptibilities, which has revealed 
that there is no singularity in the infinite volume limit and, thus, the 
transition remains an analytic crossover even at $eB=3.25\GeVt$. 

In addition, we considered the asymptotically strong magnetic field limit, 
and simulated the corresponding effective theory on the lattice.
This limiting effective theory -- an anisotropic pure gauge theory -- 
was found to exhibit a first-order deconfinement phase transition. 
Together with our findings 
above, this implies the existence of a critical point in the QCD phase 
diagram. To provide a first estimate for the magnetic field $B_{\rm CP}$ 
corresponding to the critical point, let us return to our results about the 
width $w(B)$ of the light quark susceptibilities. We have seen that 
the width is reduced as the magnetic field grows, see the right panel of 
Fig.~\ref{fig:susc_ud}. 
Assuming a linear 
dependence of $w$ on $B$ and extrapolating in the magnetic field we find that 
$w$ vanishes at
\be
eB_{\rm CP} \approx 10(2)\GeVt.
\label{eq:BCEP}
\ee
In light of the fact that the $B$-dependence of some of our 
observables (e.g.\ of $T_c$ and of $w_0$) tends to flatten out 
as $B$ grows, this first estimate should rather be taken as 
a lower bound for $B_{\rm CP}$. 
We mention that in order to simulate with magnetic fields of strengths 
comparable to that in Eq.~(\ref{eq:BCEP}), 
lattices with $N_t\gtrsim 28$ are required, out of reach for current computational 
resources.

\begin{figure}[t]
 \centering
\includegraphics[width=8.4cm]{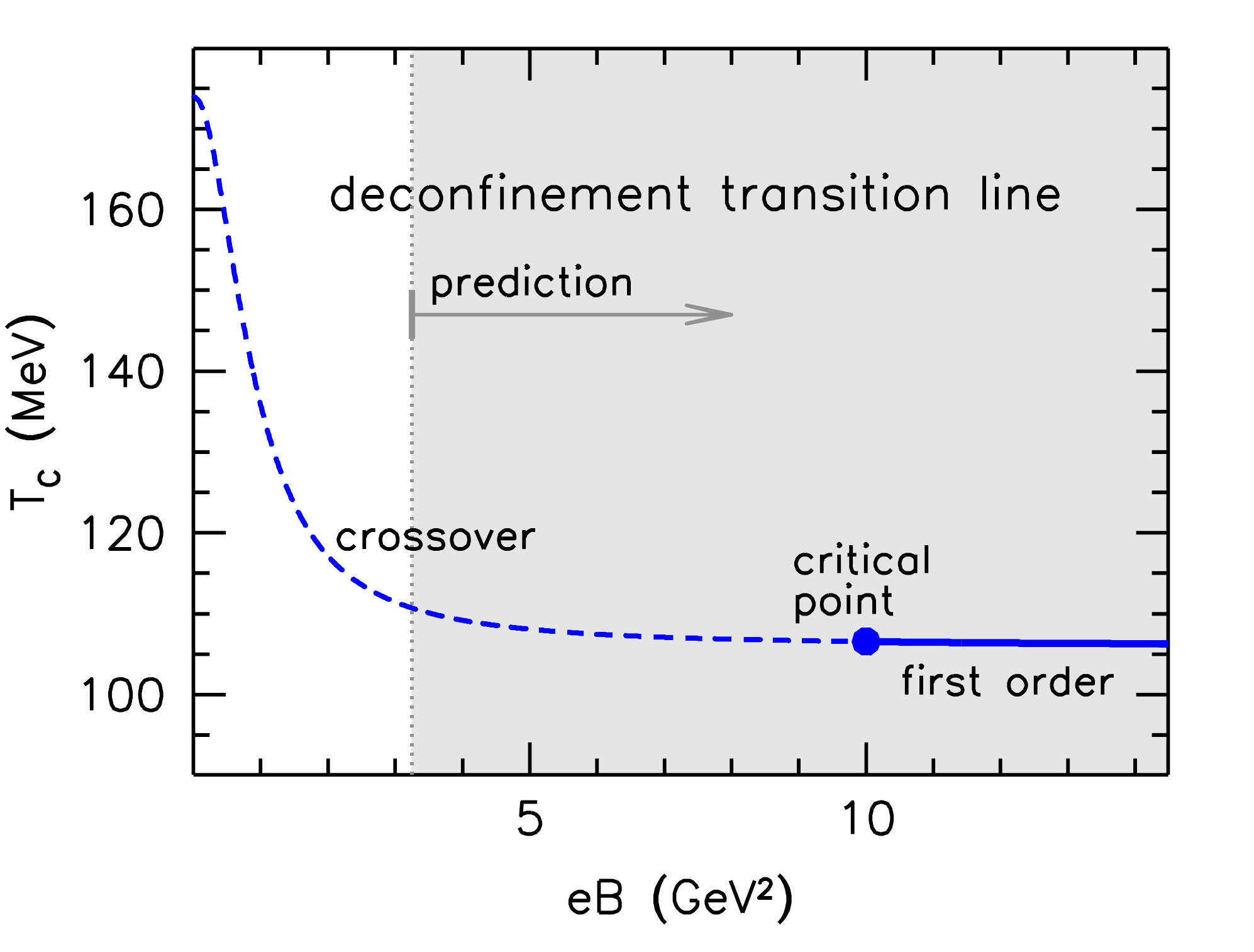}
\vspace*{-.4cm}
 \caption{\label{fig:pd_spec} The deconfinement transition temperature against the 
background magnetic field. 
The results of our full lattice QCD simulations (white background) are complemented 
by the prediction (gray background) based on the results corresponding to the 
$B\to\infty$ limit and on the extrapolation of the light quark susceptibility 
peak to high magnetic fields (see the text).}
\end{figure}

In the absence of a priori known dimensionful scales in the $B\to\infty$ 
system, we could not determine $\lim_{B\to\infty}T_c$ in physical units. 
Nevertheless, 
the deconfinement 
transition temperature of the anisotropic theory
is expected to be much 
smaller than
$T_c(B=0)$~\cite{Miransky:2002rp,*Miransky:2015ava}\footnote{
The discussion in 
Ref.~\cite{Miransky:2002rp,*Miransky:2015ava} bases on renormalization group arguments and on 
the separation of scales 
$\lambda_{\rm QCD} \ll m_d \ll \sqrt{eB}$ (here $\lambda_{\rm QCD}$ is the dynamical scale 
of the large-$B$ theory) to conclude that $\lambda_{\rm QCD}\ll \Lambda_{\rm QCD}$ 
for a very broad range of magnetic fields -- in fact, up to $eB$ being millions of orders of magnitudes 
larger than $\Lambda_{\rm QCD}^2$. 
To find the complete $B$-dependence of the running of the strong coupling
and to prove rigorously that $\lambda_{\rm QCD}\ll \Lambda_{\rm QCD}$ even in the $B\to\infty$ limit,
a full treatment of the divergent 
one-loop Feynman diagrams in the presence of background magnetic fields would be necessary. 
Without relying on the lowest-Landau-level approximation -- which might not be justified for 
the case of divergent diagrams -- this is a very difficult task.}. 
Our results for the combination $T_cw_0$ are compatible with this prediction
and suggest a gradual reduction of the deconfinement transition temperature 
as $B$ is increased\footnote{
Note that our setup at $B\to\infty$ describes the low-energy effective 
action for gluons. Thus, the results in the anisotropic theory 
have no implications for the chiral 
transition. For more details on this point, see Ref.~\cite{Miransky:2002rp,*Miransky:2015ava}. 
}.
Taking these aspects into account, Fig.~\ref{fig:pd_spec} 
represents a sketch of the deconfinement transition line in the QCD phase 
diagram for a broad range of magnetic fields.

The reader might wonder whether it is possible 
that the crossover at $eB\le 3.25 \GeVt$ and the first-order 
transition in the asymptotic limit are not connected by a single line.
To see that this is not the case, 
note that by varying the anisotropy parameter $\kappa$, 
one can continuously deform the anisotropic theory 
to usual pure gauge theory, as was demonstrated in Fig.~\ref{fig:kappa1}. 
Furthermore, the isotropic pure gauge theory can be thought of as QCD with infinitely heavy 
quarks and thus can be continuously transformed into full QCD by increasing 
the inverse quark masses from zero to their physical values. 
Thus, the transition we identified at $B\to\infty$ 
is indeed the same deconfinement transition that occurs at low magnetic fields. 

Let us highlight that according to this discussion, 
having a {\it decreasing} deconfinement transition 
temperature is actually natural to QCD. 
Furthermore, since the $B\to\infty$ limit is independent of the quark 
masses\footnote{As long as the quark masses are finite -- note that 
the $m\to\infty$ and $B\to\infty$ limits cannot be interchanged.},
a similar reduction of $T_c$ by the magnetic field should also take place in QCD 
with heavier-than-physical quarks. 
However, in the latter case this reduction most probably follows an 
initial increase in the transition temperature, cf.\ 
Refs.~\cite{D'Elia:2010nq,Bali:2011qj}.
Indeed, recent lattice results employing overlap fermions and 
pion masses of about $500\MeV$ indicate 
inverse catalysis to occur around the transition temperature at the 
magnetic field
$eB\approx1.3\GeVt$~\cite{Bornyakov:2013eya}.

Finally, we note that 
magnetic fields well above the strength~(\ref{eq:BCEP}) are predicted to be 
generated during the electroweak phase transition in the early 
universe~\cite{Vachaspati:1991nm}. If these fields remain strong enough 
until the QCD epoch, the emerging first-order phase transition might have 
several exciting consequences. 
Via supercooling, bubbles of the confined phase can be formed as the 
temperature drops below $T_c$, leading to large inhomogeneities, 
important for nucleosynthesis~\cite{Applegate:1985qt}. 
Collisions between the bubbles can also lead to 
the emission of gravitational waves and, thus, leave an imprint on the 
primordial gravitational spectrum~\cite{Witten:1984rs}. 
An absence of such signals, in turn, would imply an upper limit for the 
strength of the primordial magnetic fields.

\acknowledgments
This work was supported by the DFG (SFB/TRR 55). The author 
thanks Igor Shovkovy for valuable comments and 
Gunnar Bali, Bastian Brandt, Falk Bruckmann,
Jan Pawlowski, K\'alm\'an Szab\'o and Andreas Sch{\"a}fer 
for enlightening discussions. 

\appendix

\section{Effective action in the asymptotic magnetic field limit}
\label{app:EH}

In this appendix we demonstrate how asymptotically strong magnetic fields 
induce an anisotropy in the gluonic sector 
using an Euler-Heisenberg-type approach.
The QCD effective Lagrangian in Euclidean space-time is
\be
\L(B) = \frac{1}{2g^2}\tr G_{\mu\nu}G_{\mu\nu} + \L^q(B,G_{\mu\nu}), \quad\quad 
\L^q(B,G_{\mu\nu}) = -\sum_{f=u,d,s}\log \det \left[ \slashed{D}(q_fB,G_{\mu\nu}) + m_f \right],
\label{eq:fullaction}
\ee
and the quark determinant will be regularized using Schwinger's proper 
time formulation~\cite{Schwinger:1951nm}. 
Since the electromagnetic field exceeds all scales in the system and in particular,
$(eB)^2\gg \tr G_{\mu\nu}^2$, we
may approximate the the chromo-fields in the fermionic action to be weak. 
In addition, we assume the chromo-fields to be covariantly 
constant, $D_\mu G_{\nu\rho}=0$ to enable a fully analytical treatment of the problem. 
Given this condition, the field strength 
can be 
gauge transformed to be constant in space-time and diagonal in color space~\cite{Dunne:2004nc}, 
$G_{\mu\nu}=\textmd{diag}(G_{\mu\nu c})$ with 
the color index $c=1,2,3$.

Let us decompose the chromo-fields to chromomagnetic/chromoelectric components,
\be
\B_\parallel=G_{xy}, \quad\quad \B_\perp = \frac{G_{xz}+G_{yz}}{2}, \quad\quad
\E_\parallel = G_{zt}, \quad\quad \E_\perp = \frac{G_{xt}+G_{yt}}{2},
\label{eq:EBdef}
\ee
parallel or perpendicular to the electromagnetic field $B$. 
The leading terms in the strong $B$-expansion are
quadratic in the chromo-fields and thus, to find the coefficients of the respective 
components, it suffices to consider 
separately the effect of $B$ and $\B_\parallel$, $B$ and $\B_\perp$, $B$ and $\E_\parallel$ and
$B$ and $\E_\perp$.
The effective Lagrangian for these components for small background magnetic fields
was determined in Refs.~\cite{Bali:2013esa,Ozaki:2013sfa}. A similar calculation, 
generalized to finite temperatures and constant Polyakov loop 
backgrounds was performed in Refs.~\cite{Bruckmann:2013oba,Ozaki:2015yja}.

Let us first take the case of $B$ and $\B_\parallel$. 
For each flavor we may choose our coordinate system such that $q_fB$ is positive. 
Then, each color component experiences a total (positive) magnetic field 
$q_fB+\B_{\parallel c}$ so that
\be
\L^q(B,\B_\parallel) = \frac{1}{8\pi^2} \sum_{f,c} m_f^2\,(q_fB+\B_{\parallel c}) \int \frac{\dd s}{s^2}\, e^{-s} \,\coth \frac{(q_fB+\B_{\parallel c})s}{m_f^2}.
\label{eq:Bpar}
\ee
Since $\B_{\parallel c}$ only appears in the sum with $q_fB$, 
the effective Lagrangian becomes independent of the chromomagnetic field 
in the limit $q_fB\gg \B_{\parallel c}$. This implies that quarks become 
insensitive to $\B_{\parallel}$, i.e.\ decouple from this gluonic component.

Next we take the case with $B$ and $\B_\perp$. Rotating our coordinate axes for 
each color component such 
that the $z$ axis points in the direction of the total magnetic field we get
\be
\L^q(B,\B_\perp) = \frac{1}{8\pi^2} \sum_{f,c} m_f^2 \sqrt{(q_fB)^2+\B_{\perp c}^2} \int \frac{\dd s}{s^2}\, e^{-s} \,\coth \frac{\sqrt{(q_fB)^2+\B_{\perp c}^2}s}{m_f^2}.
\label{eq:Bperp}
\ee
In the strong $B$ limit, this becomes independent of $\B_{\perp c}$, signaling 
that quarks decouple from the perpendicular chromomagnetic component of the gluons as well.

For a perpendicular chromoelectric field, for each color component 
we can perform the Lorentz 
transformation that eliminates the electric field and, in turn, gives 
a total magnetic field $\sqrt{(q_fB)^2+\E_{\perp c}^2}$. The corresponding 
effective Lagrangian equals Eq.~(\ref{eq:Bperp}), but 
with $\B_{\perp c}$ replaced by $\E_{\perp c}$. This implies the decoupling 
of quarks from the perpendicular chromoelectric fields.

Finally, for a parallel chromoelectric field $\E_\parallel$, 
we have a Landau problem in the $x-y$ as well as 
in the $z-t$ planes, giving 
\be
\L^q(B,\E_\parallel) = \frac{1}{8\pi^2} \sum_{f,c} q_fB \,\E_{\parallel c} \int \frac{\dd s}{s}\, e^{-s} \,\coth \frac{q_fBs}{m_f^2}\, \coth \frac{\E_{\parallel c} s}{m_f^2}.
\label{eq:Epar}
\ee
Taking the limit $q_fB\gg \E_{\parallel c},m_f^2$, we see that -- unlike for the other components above 
-- a non-trivial dependence 
on $\E_\parallel$ remains. 
We are interested in the quadratic term, proportional to $\tr \E_{\parallel}^2$, which 
contributes\footnote{
Here we omitted a divergent term of the form 
$\tr \E_\parallel^2 \log \Lambda/m_f$, 
where $\Lambda$ is a cutoff entering as the lower endpoint of the proper time 
integration $s_0\propto 1/\Lambda^2$. 
This divergence can be eliminated by
the multiplicative renormalization of the wave function $\E_\parallel$ and of the gauge coupling 
$g$~\cite{Schwinger:1951nm}. 
Closer inspection of Eqs.~(\ref{eq:Bpar}) and~(\ref{eq:Bperp}) shows that the same type of divergence
is present for the other components as well. Thus, these $B$-independent terms merely 
represent an isotropic redefinition 
of the gauge coupling $g$, which does not alter the form of the effective Lagrangian for 
strong magnetic fields.
Another divergence, independent of the gluonic field strengths, takes the form 
$(q_fB)^2 \log \Lambda/m_f$ and is canceled by the renormalization of $B$ and of 
$q_f$~\cite{Schwinger:1951nm}. Thus, the necessary renormalizations at $B\to\infty$ 
are of the same type as for the theory at small magnetic fields.
} 
to the gluonic Lagrangian $\tr G_{\mu\nu}^2$ of Eq.~(\ref{eq:fullaction}),
\be
\L^q(B,\mathcal{O}(\E_{\parallel}^2)) = \frac{1}{24\pi^2} \sum_{f,c} q_fB \, \frac{\E_{\parallel c}^2}{m_f^2} \int \dd s \, e^{-s} = \kappa(B)\, \tr \E_{\parallel}^2,
\quad\quad\quad
\kappa(B) \equiv \frac{1}{24\pi^2} \sum_f |q_f/e| \frac{|eB|}{m_f^2}.
\label{eq:kappa0}
\ee
Altogether, the asymptotically strong magnetic field limit of the QCD effective 
Lagrangian indeed equals Eq.~(\ref{eq:LagB}). 
Thus we find that the chromo-dielectric constant is
enhanced in the direction of the background magnetic field,  
and the coefficient $\kappa(B)$ coincides with the result of Ref.~\cite{Miransky:2002rp,*Miransky:2015ava}. 

\begin{wrapfigure}{r}{9cm}
 \centering
 \vspace*{-.3cm}
 \includegraphics[width=8.7cm]{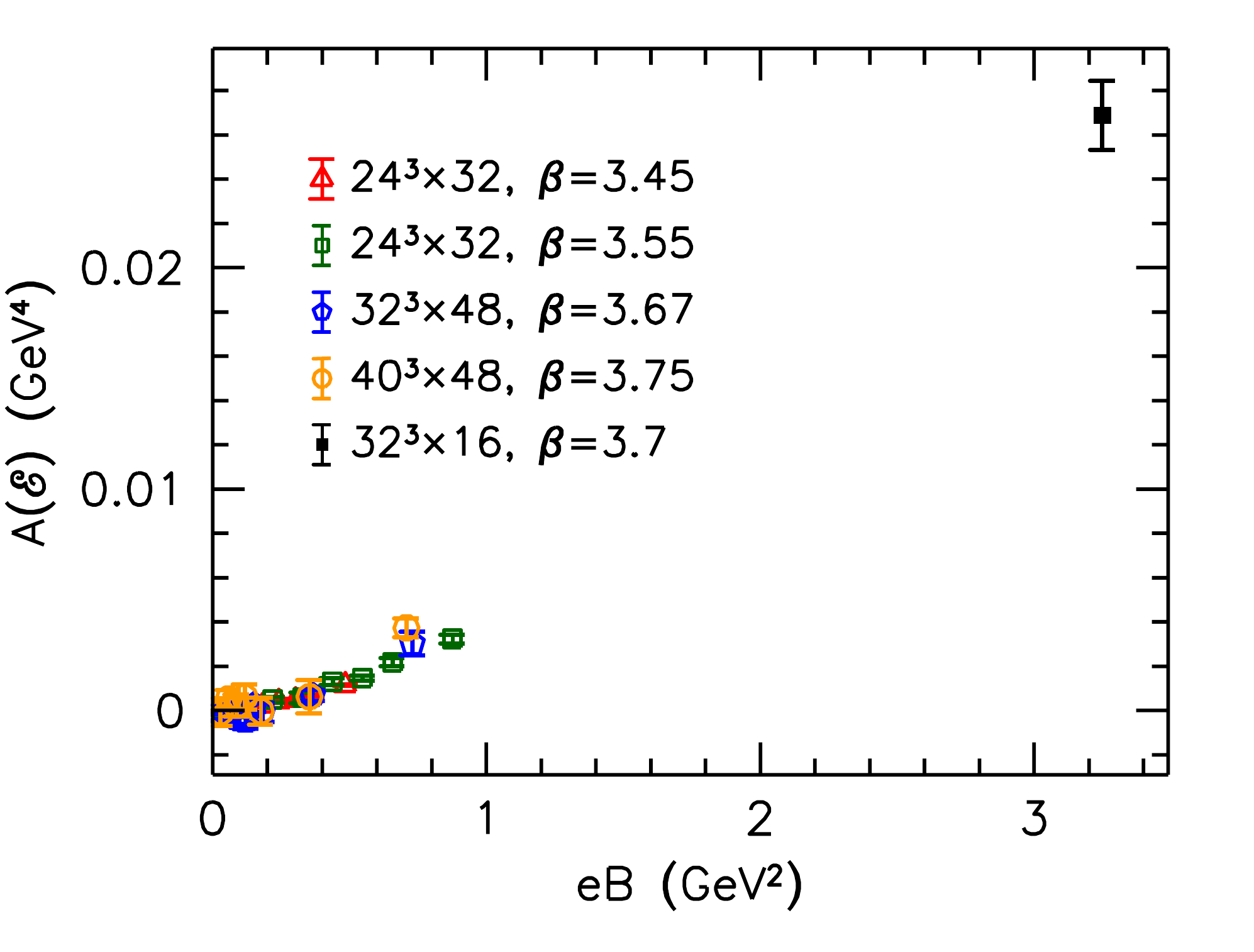}
 \vspace*{-.2cm}
 \caption{\label{fig:aniso}
 The anisotropy of the chromoelectric gluonic field strength component in full QCD.
 }
\end{wrapfigure}
Having $\kappa\gg1$ in the action implies that the corresponding 
gluonic field strength component $\tr\E_\parallel^2$ is strongly suppressed. 
In other words, the anisotropy in the chromoelectric part of the action density,
\be
A(\E)=\frac{1}{V_4} \frac{1}{g^2} \expv{\tr\E_\perp^2 - \tr\E_\parallel^2},
\ee
is enhanced by $B$.
To back up this prediction, in Fig.~\ref{fig:aniso} we plot $A(\E)$ 
as a function of the magnetic field, based on our zero-temperature 
results at $eB<1\GeVt$~\cite{Bali:2013esa} 
and the measurements at $eB=3.25\GeVt$ at our lowest temperature $T\approx 75\MeV$. 
The results clearly indicate that the anisotropy is positive and strongly 
increased as $B$ grows.
We note that due to the coupling between the gluonic field strength components, 
a similar anisotropy in the chromomagnetic sector also appears, altogether 
giving rise to the hierarchy $\tr\B_\parallel^2>\tr\B_\perp^2=\tr\E_\perp^2>\tr\E_\parallel^2$ 
at low temperatures. 
The same hierarchy is also observed in the anisotropic gauge theory\footnote{
To see how the anisotropic di{\it electric} constant affects the chromo{\it magnetic} 
components, it is instructive to consider the gauge potential $A_\mu$. A large value 
of $\kappa$ implies a suppression of $\tr\E_\parallel^2$ and a corresponding suppression 
of the fluctuations in $A_z$ and in $A_t$. This suppression propagates into the 
magnetic sector and creates the anisotropy between $\tr\B_\perp^2$ and 
$\tr\B_\parallel^2$. Indeed, while the former contains $A_z$, the latter does not.}.

We note that the 
calculation leading to Eq.~(\ref{eq:LagB}) 
can also be performed for nonzero temperatures. 
At $T>0$ an additional factor appears in the proper time integral due to 
the sum over Matsubara frequencies, containing an elliptic $\Theta$-function. 
This factor decouples from the $B$-dependence, implying
that even for $T>0$, only the chromo-dielectric constant is affected.
The coefficient $\kappa$ is, however, altered as
\be
\kappa(B,T) = \frac{1}{24\pi^2} \sum_f |q_f/e| \frac{|eB|}{m_f^2} \int
\dd s \,e^{-s}\, \Theta_3\Big[\frac{\pi}{2},e^{-m_f^2/(4sT^2)}\Big].
\ee
The integral over $s$ equals unity at $T=0$ and is reduced monotonously 
(and smoothly) as the temperature 
grows. Simulating the anisotropic gauge theory according to the 
Lagrangian~(\ref{eq:LagB}) on the lattice, we found that the 
theory exhibits a first-order phase transition. 
Thus, since the smooth $\kappa(T)$ dependence does not affect the discontinuous 
transition, in order to locate the critical temperature 
it suffices to simulate the theory at fixed (large) $\kappa$ values. 

\section{Simulating anisotropic pure gauge theory on the lattice}
\label{sec:aniso_sim}

In this appendix we discuss the simulation algorithm for the anisotropic pure 
gauge theory described by the Lagrangian~(\ref{eq:LagB}).
The corresponding path integral
\be
\Z = \int \D U\, e^{-\beta S_g^{\rm aniso}},
\ee
can be simulated directly on the lattice. 
Here, $U$ denotes the gauge links, $\beta=6/g^2$ 
is the inverse gauge coupling and the anisotropic gauge action reads
\be
S_g^{\rm aniso} = \sum_{\mu<\nu} \frac{1}{3}\,\textmd{Re} \,\tr P_{\mu\nu} \cdot\kappa_{\mu\nu}, \quad\quad\quad
\kappa_{\mu\nu}=\begin{cases}
             1+\kappa(B)/\beta, &\quad \mu=z, \nu=t, \\
             1, &\quad \textmd{otherwise},
            \end{cases}
\label{eq:Sganiso}
\ee
where $P_{\mu\nu}$ are linear combinations of closed loops lying in the $\mu-\nu$ 
plane. We take the tree-level Symanzik improved gauge action such that these loops 
include the $1\times1$ plaquettes $U_{\mu\nu}^{1\times 1}$ and the $2\times1$ rectangles $U_{\mu\nu}^{2\times 1}$ with appropriately 
tuned coefficients~\cite{Weisz:1982zw},
\be
\label{eq:weisz}
P_{\mu\nu} = -\frac{1}{12} \left(\mathds{1}-U_{\mu\nu}^{2\times 1}\right) + \frac{5}{3} \left(\mathds{1}-U_{\mu\nu}^{1\times 1}\right).
\ee
The correspondence between the continuum and lattice expressions reads
\be
\tr\E_\parallel^2 =  2 \,\textmd{Re} \,\tr P_{tz},
\quad\quad
\tr\E_\perp^2 =  \textmd{Re} \,\tr [P_{tx} + P_{ty} ],
\quad\quad
\tr\B_\parallel^2 =  2 \,\textmd{Re} \, \tr P_{xy},
\quad\quad
\tr\B_\perp^2 = \textmd{Re} \, \tr [ P_{yz} + P_{xz} ].
\label{eq:Pcorr}
\ee

To simulate this theory, we use an overrelaxation/heatbath algorithm, 
based on the isotropic pure gauge implementation by the MILC collaboration~\cite{milc7.6}. 
One trajectory consists of one 
overrelaxation step followed by four heatbath steps. 
The simulation at finite anisotropy coefficient $\kappa$ simply involves 
multiplying the plaquettes and rectangles lying in the $z-t$ plane by $\kappa$. 
We observe that autocorrelation times grow large as $\kappa$ increases, similarly 
to the issue of critical slowing down of the isotropic theory at large $\beta$. 
This prohibits approaching $\kappa\to\infty$, necessary for the asymptotically strong 
magnetic field limit. 
However, it is possible to modify the algorithm to simulate directly at $\kappa=\infty$. 
In this limit, the $z-t$ component of the action and the remaining five components
decouple, 
and the links are restricted to the subspace $\Omega[U]$ of configurations, 
where $P_{zt}$ is minimal. This subspace is defined by
\be
\Omega[U] = \{U_\mu \,|\,U_{zt}^{1\times1}=\mathds{1} \}.
\label{eq:VU}
\ee
Indeed, any fluctuation in the link variables that leads off of this subspace 
makes the action infinitely large and is thus forbidden.
(Note that if $U_{zt}^{1\times1}$ 
equals the unit matrix, then so does $U_{zt}^{2\times1}$.)

\begin{figure}[t]
 \centering
 \mbox{
 \includegraphics[width=6cm]{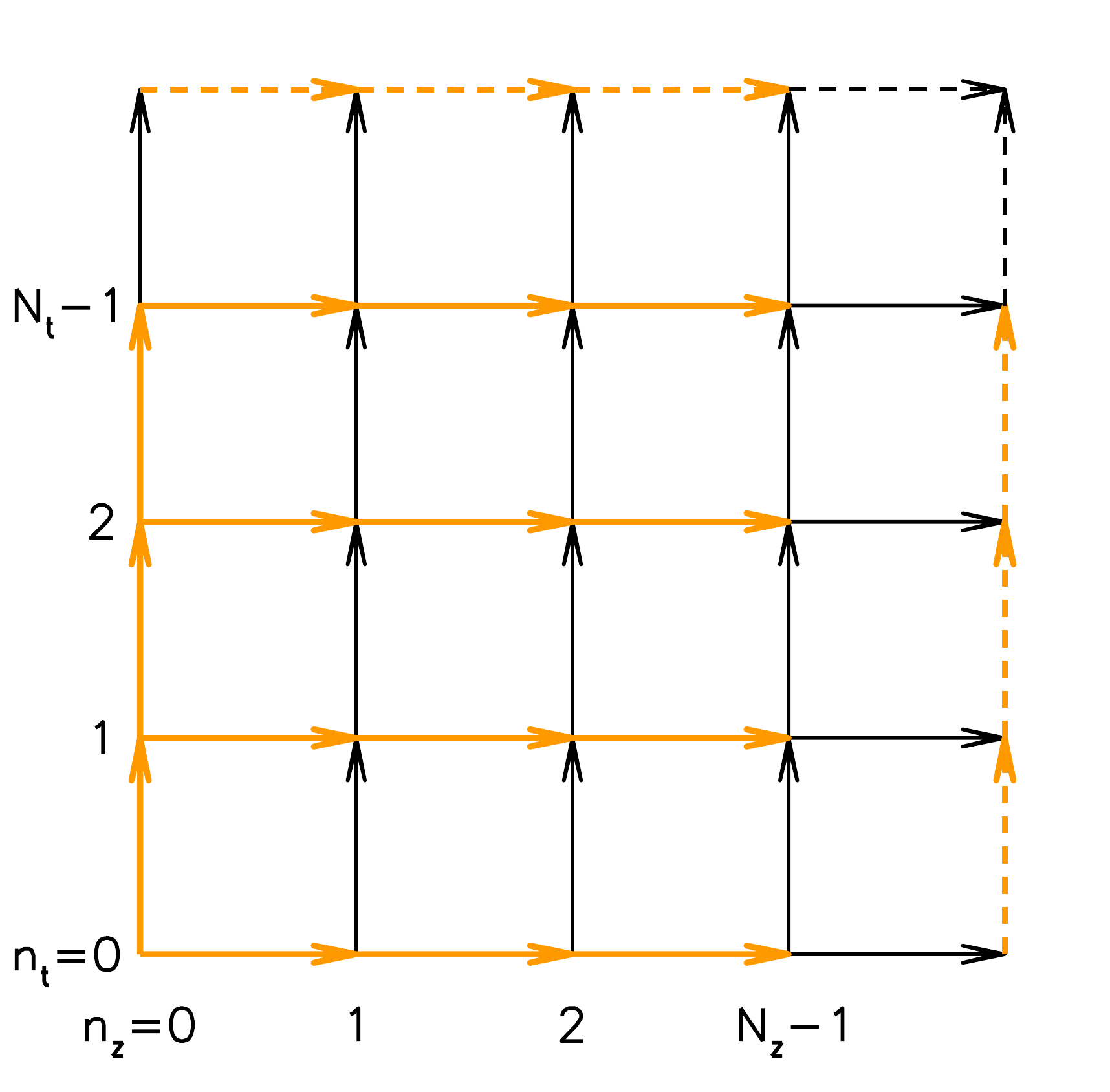} \;
 \includegraphics[width=6cm]{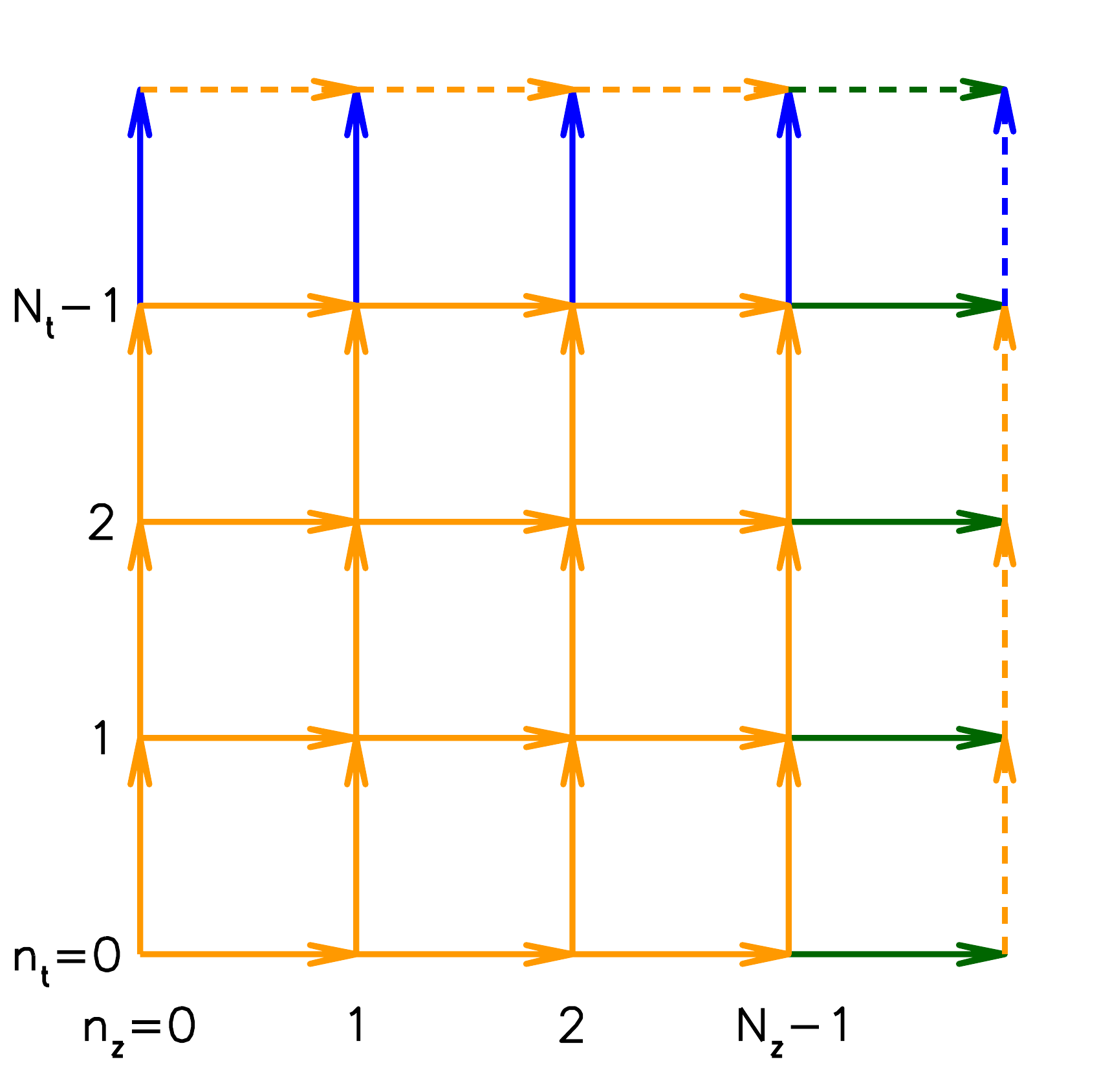}}
 \caption{\label{fig:maxtree} One $z-t$ plane of the lattice with periodic boundary 
 conditions in both directions. The dashed arrows at $n_z=N_z$ ($n_t=N_t$) 
 indicate the copies of the links at $n_z=0$ ($n_t=0$). 
 Left panel: the gauge links on the maximal tree (yellow arrows) are fixed to the unit 
 matrix. Right panel: in order for the configuration to belong to the 
 subspace~\protect(\ref{eq:subspace}), 
 further gauge links are set to unity (yellow) and the 
 green and blue links are set equal, respectively. Finally, the blue and green 
 `degenerate' Polyakov loops must commute with each other.}
\end{figure}

We thus have to parameterize the subspace $\Omega[U]$ in terms of the gauge links $U_\mu$. 
Let us label the lattice sites by $n=(n_x,n_y,n_z,n_t)$ with $0\le n_\mu<N_\mu$. 
To find the parameterization of Eq.~(\ref{eq:VU}), it is advantageous to fix the links to $\mathds{1}$ 
on a so-called maximal tree. The specific choice for the tree is shown in the left panel of 
Fig.~\ref{fig:maxtree}. (Note that Faddeev-Popov fields are absent for such a gauge fixing~\cite{Montvay:1994cy}.) 
In order to have unit plaquettes for $n_z<N_z-1$ and 
$n_t <N_t-1$, 
all $t$-links must be set to unity at these sites. To have unit plaquettes on the last 
$z$-slice, 
all $z$-links at $N_z-1$ must be set equal, denoted by $L_z$. Similarly, all the $t$-links at 
$N_t-1$ must be set equal, denoted by $L_t$, see the visualization in the right panel of 
Fig.~\ref{fig:maxtree}. These remaining links correspond to the local Polyakov loops in the $z$- 
and in the $t$-direction, lying in the $z-t$ plane at a given $n_x$ and $n_y$. 
Finally, to ensure that the plaquette at the corner $n_z=N_z-1$, $n_t=N_t-1$ is unity, we need 
$L_zL_t L_z^\dagger L_t^\dagger = \mathds{1}$, i.e.\ $L_z$ and $L_t$ must commute. 
Altogether, the subspace in question reads
\be
\begin{split}
\Omega[U] = \{ U_\mu \,|\, &U_z(n)=\mathds{1} \;\forall n_z\neq N_z-1, \\
&U_t(n)=\mathds{1} \;\forall n_t\neq N_t-1, \\
&U_z(n_x,n_y,N_z-1,n_t)=L_z(n_x,n_y) \;\forall n_t,\\
&U_t(n_x,n_y,n_z,N_t-1)=L_t(n_x,n_y) \;\forall n_z, \\
&[L_z(n_x,n_y), L_t(n_x,n_y)] = 0 \}.
\end{split}
\label{eq:subspace}
\ee
Notice that the `degenerate' timelike Polyakov loop $L_t$ 
(represented by the blue arrows in the right panel of Fig.~\ref{fig:maxtree})
appears multiple times in the action -- in fact, in $4N_z$ 
plaquettes and in $12N_z$ rectangles. 
The corresponding 
`staples' are all taken into account in the update of $L_t$ (and similarly for $L_z$).

There are several ways to fulfill the commutativity relation $[L_z,L_t]=0$. 
One possibility (setup A) is to simply set $L_z=\mathds{1}$ for all $n_x$ and $n_y$. 
Another approach (setup B) is to constrain $L_z$ to be a center element, 
$L_z(n_x,n_y)\in\mathds{Z}_3$. 
The two setups only differ on a set whose measure vanishes in the limit, where all lattice 
extents are taken to infinity. 
Note that the expectation value of the average $z$-Polyakov loop $P^{(z)}$
[defined similarly as the usual Polyakov loop $P$, Eq.~(\ref{eq:ploop})] is three for setup A, whereas 
it is zero for setup B, if the spatial size of the system is large enough. 
Nevertheless, we checked that observables sensitive to the finite temperature 
transition ($P$, the gauge action, etc.) all have vanishing 
correlators with $P^{(z)}$. 
In fact, we found that the setups A and B give identical results
for $P$ and for the gauge action for 
all values of the inverse gauge coupling $\beta$ on the $16^3\times 4$ lattices. 
In other words, center symmetry breaking in the $z$ direction appears to be 
completely irrelevant for the deconfinement 
phase transition. 

In addition, we also tried allowing both $L_z$ and $L_t$ to be general $\mathrm{SU}(3)$
matrices (which violates the commutativity relation). This approach (setup C) 
turned out to introduce negligible
differences in the results\footnote{
The most general prescription is obtained by gauge transforming $L_t$ to
diagonal form. The commutativity relation then ensures that also $L_z$ is
diagonal. In the infinite lattice size limit, this setup ought to give identical 
results as well. We leave this check for a future study.}. 
For the simulations presented in the body of the paper, we considered setup A
and set $L_z=\mathds{1}$ throughout the lattice. 
As a consistency check, besides the overrelaxation/heatbath 
algorithm, we also considered a hybrid Monte-Carlo update and found that the two 
give fully consistent results. 

\bibliographystyle{JHEP_mcite}
\bibliography{largeB}

\end{document}